\documentclass[useAMS,usenatbib]{mn2e}
\usepackage{hyperref}
\usepackage{graphicx}
\usepackage{multirow}
\usepackage{times}

\def \deg{$^\circ$}

\title{Exploring the role of the Sun's motion in terrestrial comet impacts}
\author[F.\ Feng and C.A.L.\ Bailer-Jones]{F.\ Feng\thanks{E-mail:ffeng@mpia.de} and C.A.L.\ Bailer-Jones\\
Max Planck Institute for Astronomy, K\"onigstuhl 17, 69117 Heidelberg
}

\begin{document}
\date{\today}

\pagerange{\pageref{firstpage}--\pageref{lastpage}} \pubyear{2014}

\maketitle

\label{firstpage}

\begin{abstract}
The cratering record on the Earth and Moon shows that our planet has been exposed to high velocity impacts for much or all of its existence. Some of these craters were produced by the impact of long period comets (LPCs).  These probably originated in the Oort cloud, and were put into their present orbits through gravitational perturbations arising from the Galactic tide and stellar encounters, both of which are modulated by the solar motion about the Galaxy. Here we construct dynamical models of these mechanisms in order to predict the time-varying impact rate of LPCs and the angular distribution of their perihelia (which is observed to be non-uniform). Comparing the predictions of these dynamical models with other models, we conclude that cometary impacts induced by the solar motion contribute only a small fraction of terrestrial impact craters over the past 250\,Myr. Over this time scale the apparent cratering rate is dominated by a secular increase towards the present, which might be the result of the disruption of a large asteroid. Our dynamical models, together with the solar apex motion, predict a non-uniform angular distribution of the perihelia, without needing to invoke the existence of a massive body in the outer Oort cloud. Our results are reasonably robust to changes in the parameters of the Galaxy model, Oort cloud, and stellar encounters.
\end{abstract}
\begin{keywords}
  Earth --- Galaxy: kinematics and dynamics --- methods: statistical
  --- solar-terrestrial relations --- comets: general --- Oort Cloud
\end{keywords}

\section{Introduction}\label{sec:introduction}

\subsection{Background}\label{sec:background} 

Comet or asteroid impacts on the Earth are potentially catastrophic
events which could have a fundamental effect on life on Earth.  While
at least one extinction event and associated crater is well documented
-- the K-T impact from 65\,Myr ago and the Chicxulub crater
\citep{alvarez80, 1991Geo....19..867H} -- a clear connection between
other craters and extinction events is less well
established. Nonetheless, we know of around 200 large impact craters
on the Earth, and doubtless the craters of many other impacts have either
since eroded or are yet to be discovered.

Many studies in the past have attempted to identify patterns in the
temporal distribution of craters and/or mass extinction events.
Some claim there to be a periodic component in the data (e.g.\
\citealp{alvarez84,raup84,rohde05,melott11}), although the reliability of
these analyses is debated, and other studies have come to other
conclusions (e.g.\ \citealp{1996EM&P...72..357G}
\citealp{1996MNRAS.279..727Y} \citealp{2000A&A...353..409J}
\citealp{bailer-jones09} \citealp{bailer-jones11,feng13}).

Of particular interest is whether these impacts are entirely random,
or whether there are one or two dominant mechanisms which account for
much of their temporal distribution. Such mechanisms need not be
deterministic: stochastic models show characteristic distributions in
their time series or frequency spectra (e.g.\
\citep{2012A&A...546A..89B}). We are therefore interested in
accounting not for the times of individual impacts, but for the impact
rate as a function of time.

In doing this we should distinguish between asteroid and comet
impacts.  Having smaller relative velocities, asteroid impacts are
generally less energetic. Asteroids originate from within a few AU of
the Sun, so their impact rate is probably not affected much by events
external to the solar system. Comets, on the other hand, originate
from the Oort cloud \citep{oort50}, and so can be affected by the
Galactic environment around the Sun.

As the solar system orbits the Galaxy, it experiences gravitational perturbations from the Galactic tide and from encountering with individual passing stars. These perturbations are strong enough to modify the orbits of Oort cloud comets to inject them into the inner solar system \citep{wickramasinghe08,gardner11}.  The strength of these perturbations is dependent upon the local stellar density, so the orbital motion of the Sun will modulate these influences and thus the rate of comet injection and impact to some degree (e.g.\ \citet{brasser10,kaib11,levison10}). As the Sun shows a (quasi)-periodic motion perpendicular to the Galactic plane, and assuming that the local stellar density varies in the same way, it has been argued that this could explain a (supposed) periodic signal in the cratering record.  Here we will investigate the connection between the solar motion and the large impact craters (i.e.\ those generated by high energy impacts) more explicitly.  We do this by constructing a dynamical model of the Sun's orbit, the gravitational potential, and the resulting perturbation of comet orbits, from which we will make probabilistic predictions of the time variability of the comet impact rate.

The dates of impact craters are not the only relevant observational
evidence available.  We also know the orbits of numerous long-period
comets (LPCs).  The orbits of dynamically new LPCs -- those which
enter into the inner solar system for the first time -- record the
angular distribution of the cometary flux.  This distribution of their
perihelia is found to be anisotropic. Some studies interpret this as
an imprint of the origination of comets \citet{bogart82, khanna83},
while others believe it results from a perturbation of the Oort Cloud.
Under this perturbation scenario, it has been shown that the Galactic
tide can (only) deplete the pole and equatorial region of the Oort
Cloud \citep{delsemme87} in the Galactic frame, and so cannot account
for all the observed anisotropy in the LPC perihelia. It has been
suggested that the remainder is generated from the perturbation of
either a massive body in the Oort Cloud \citep{matese99, matese11} or
stellar encounters \citet{biermann83,dybczynski02}.

\subsection{Overview}\label{sec:overview} 

Assuming a common origin of both the large terrestrial impact craters
and the LPCs, we will construct dynamical models of the flux and
orbits of injected comets as a function of time based on the solar
motion around the Galaxy. Our approach differs from previous work in
that we (1) simulate the comet flux injected by the Galactic tide and
stellar encounters as they are modulated by the solar motion; (2) use an
accurate numerical method rather than averaged Hamiltonian
\citep{fouchard04} or Impulse Approximation
\citep{oort50,rickman76,rickman05} in the simulation of cometary
orbits; (3) take into account the influence from the Galactic bar and
spiral arms; (4) test the sensitivity of the resulting cometary flux
to varying both the initial conditions of the Sun and the parameters
of the Galaxy potential, Oort Cloud, and stellar encounters.

We build the dynamical models as follows. Adopting models of the
Galactic potential, Oort Cloud and stellar encounters, we integrate
the cometary orbits in the framework of the AMUSE software
environment, developed for performing various kinds of astrophysical
simulations \citep{portegies13, pelupessy13}. The
cometary orbits can be integrated with the perturbation from either
the Galactic tide, or stellar encounters, or both. All three are investigated. In principle, we can
build a three-parameter dynamical model for the variation of the
impacting comet flux as a function of time, Galactic latitude, and
Galactic longitude. In practice we reduce this three-parameter model
to a 1-parameter model of the variation of the comet impact rate over
time, and a 2-parameter model of the angular distribution of the
perihelia of LPCs.  A further simplification is achieved by replacing
the full numerical computations of the perturbations by separating
proxies for the tide-induced comet flux and for the encounter-induced
comet flux.  These are shown to be good approximations which accelerate
considerably the computations.

We combine the predictions of the comet impact history with a
(parameterized) component which accounts for the crater preservation
bias (i.e.\ older craters are less likely to be discovered) and the asteroid impact rate.
We then use Bayesian model comparison to compare the predictions of
this model over different ranges of the model parameters to the
observed cratering data, using the crater data and statistical method
presented in \citet{bailer-jones11}.

We obtain the 2-parameter model for the angular distribution of the
perihelia of LPCs by integrating the full 3-parameter model over time.
Because we no longer need the time resolution, we actually perform a
separate set of numerical simulations to build this model.  We then
compare our results with data on 102 new comets with accurately
determined semi-major axes (the ``class 1A'' comets of
\citet{marsden08}).

This paper is organized as follows. We introduce, in section
\ref{sec:data}, the data on the craters and LPCs. In section
\ref{sec:simulation} we define our models for
the Galactic potential, the Oort cloud, and for stellar encounters,
and describe the method for the dynamical simulation
of the comet orbits.  In section \ref{sec:bayesian} we summarize
the Bayesian method of model comparison. In
section \ref{sec:impact} we use the dynamical model to 
construct the 1-parameter model of the
cometary impact history. In Section
\ref{sec:comparison}, we compare our dynamical time series models of
the impact history with other models, to assess how well the data
support each. In section \ref{sec:ADP} we use the dynamical model again, but
this time to predict the distribution of the perihelia of LPCs (the 2-parameter
model), which we compare with the data. A test of the sensitivity of these
model comparison results to the model parameters is made in section
\ref{sec:sensitivity}. We discuss our results and conclude in section
\ref{sec:conclusion}.

The main symbols and acronyms used in this article are summarized
in Table \ref{tab:symbol}.
\begin{table}
\centering
\caption{Glossary of main acronyms and variables}
\label{tab:symbol}
\begin{tabular}{l l}
  \hline
  \hline
  Symbol & Definition\\
  \hline
  PDF       & probability density function \\
  LSR       & local standard of rest \\
  HRF       & heliocentric rest frame \\
  BP        & before present\\
  LPC       & long-period comet\\
  $s_j$      & crater age\\
  $\sigma_t$ & age uncertainty of crater   \\
  $s^{up}$    & upper limit of the age of crater\\
  ${\vec r}_{\rm enc}$  &impact parameter or perihelion of encounter\\
  $ \vec{v}_\star$&velocity of a star in the LSR\\
  $ \vec{v}_{\rm enc}$&velocity of the stellar encounter relative to the Sun\\
  $ b_\star $ &Galactic latitude of $\vec{v}_\star$   \\
  $ l_\star $ &Galactic longitude of $\vec{v}_\star$   \\
  $ b_{\rm enc} $ & Galactic latitude of $\vec{v}_{\rm enc}$   \\
  $ l_{\rm enc} $ &Galactic longitude of $\vec{v}_{\rm enc}$   \\
  $ b_p$ & Galactic latitude of the perihelion of a stellar encounter \\
  $ l_p$ & Galactic longitude of the perihelion of a stellar encounter \\
  $ b_c $    & Galactic latitude of cometary perihelion \\
  $ l_c $    &Galactic longitude of cometary perihelion\\
  $ q $      & perihelion distance\\
  $ a $      & semi-major axis\\
  $ e $      & eccentricity\\
  $M_{\rm enc}$   &mass of a stellar encounter\\
  $v_{\rm enc}$ &  speed of a star at encounter\\
  $r_{\rm enc}$ & distance of a star at encounter\\
  $f_c$     &injected comet flux relative to the total number of comets\\
  $\bar{f}_c$     &averaged $f_c$ over a time scale\\
  $\gamma$   &parameter of impact intensity$\frac{M_{\rm enc}}{v_{\rm enc}r_{\rm enc}}$\\
  $\gamma_{\rm bin}$ &normalized maximum $\gamma$ in a time bin\\
  $G_1$, $G_2$&coefficients of radial tidal force\\ 
  $G_3$     &coefficient of vertical tidal force\\ 
  $\rho$    &stellar density\\
  $\eta$    & ratio between the trend component and $f_c$\\
  $\xi$     & ratio between the tide-induced flux and encounter-induced flux\\
  $\kappa$  & angle between ${\vec r}_{\rm enc}$ and the solar apex\\
  $M_s$ & mass of the Sun\\
  \hline
\end{tabular}
\end{table}

\section{Data}\label{sec:data} 

\subsection{Terrestrial craters}\label{sec:craters}

\begin{figure}
\centering
\includegraphics[scale=0.42]{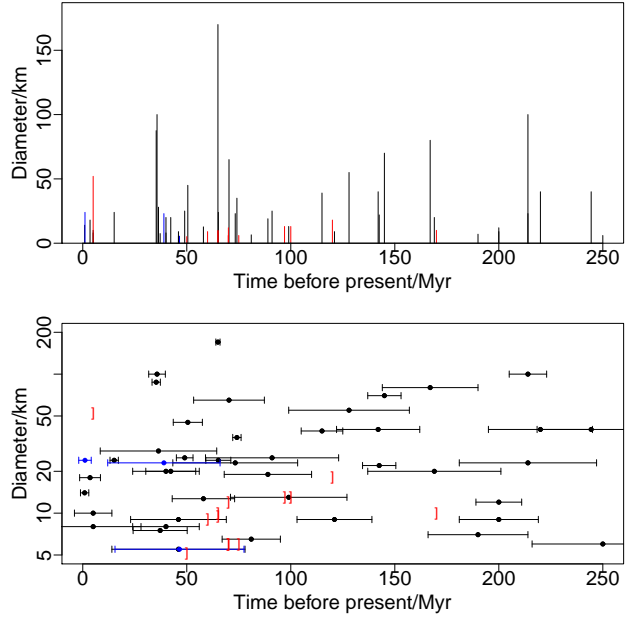}
\caption{The diameters and ages of the 59 craters with (bottom) and
  without (top) age uncertainties plotted. The blue points/lines
  indicate the craters with assigned age uncertainties. The red
  lines/brackets indicate the upper ages of the craters without
  well-defined ages. Adapted from \citet{bailer-jones11}. }
\label{fig:set3lim}
\end{figure}

The data of craters we use in this work is from the {\em Earth Impact
  Database} (EID) maintained by the Planetary and Space Science Center
at the University of New Brunswick.  We restrict our analysis to
craters with diameter $>5$ km and age $<250$\,Myr in order to reduce
the influence of crater erosion (although an erosion effect is
included in our time series models).  We select the following data
sets defined by \citet{bailer-jones11}
\begin{itemize}
\item{\bf basic150} (32 craters) age $\le$ 150\,Myr, $\sigma_t$ original
\item{\bf ext150} (36 craters) age $\le$ 150\,Myr, original or assigned
\item{\bf full150} (48 craters) ext150 plus craters with $s^{up}\le$ 150\,Myr
\item{\bf basic250} (42 craters) age $\le$ 250\,Myr, $\sigma_t$ original
\item{\bf ext250} (46 craters) age $\le$ 250\,Myr, original or assigned
\item{\bf full250} (59 craters) ext250 plus craters with $s^{up}\le$ 250\,Myr
\end{itemize}
The terms ``basic'', ``ext'', and ``full'' refer to the inclusion of
craters with different kinds of age uncertainties.  ``original
$\sigma_t$'' means that just craters with measured age uncertainties
are included. ``original or assigned'' adds to this craters for which
uncertainties have been estimated. The ``full'' data sets further
include craters with just upper age limits (\citealp{bailer-jones11}
explains how these can be used effectively).  As the size
of the existing craters is determined by many factors, e.g.\ the
inclination, velocity and size of the impactor, the impact surface,
and erosion, we only use the time of occurrence ($s_j$) of each impact
crater and its uncertainty ($\sigma_j$). Figure \ref{fig:set3lim}
plots the size and age of the 59 craters we use in the model comparison in Section
\ref{sec:comparison}.

\subsection{Long-period comets}\label{sec:LPCs}

\begin{figure}
  \centering
  \includegraphics[scale=0.41]{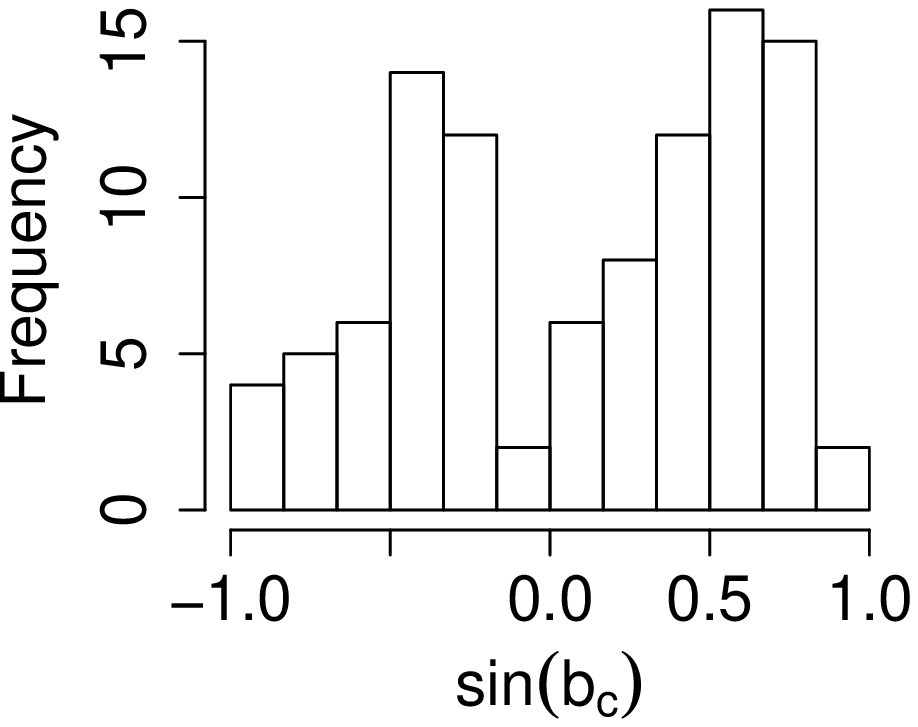}
  \includegraphics[scale=0.41]{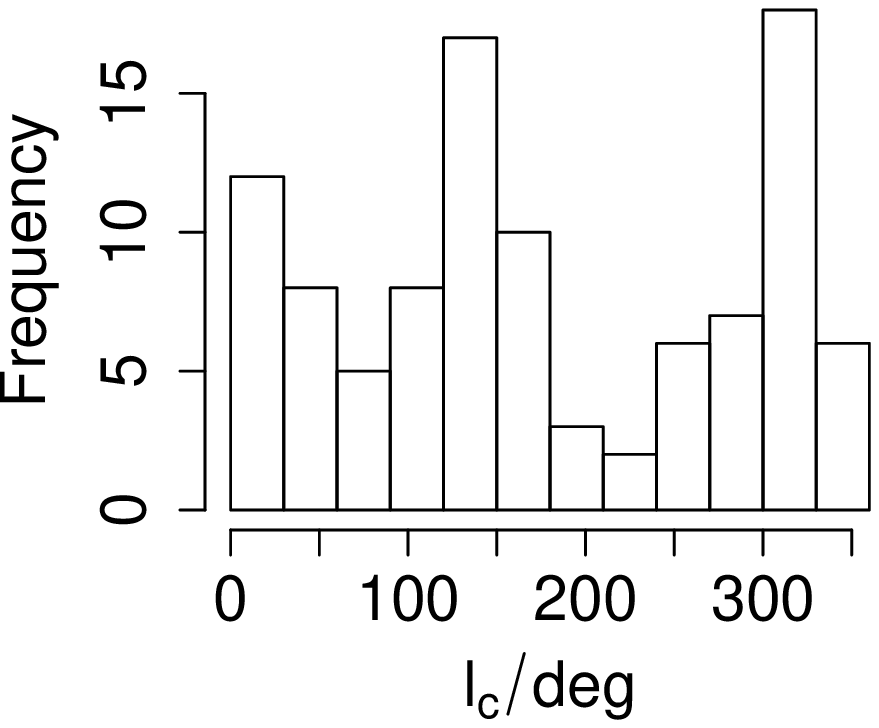}
  \caption{The distribution of $\sin b_c$ (left panel) and $l_c$ (right panel) of perihelia of the 102 LPCs. }
\label{fig:LPC1A_bl}
\end{figure}

The LPCs we use are the 102 dynamically new comets (i.e. class 1A) identified by \citet{marsden08} and discussed by \citet{matese11}. 
Figure \ref{fig:LPC1A_bl} shows the distribution over the Galactic 
latitude ($b_c$) and longitude ($l_c$) of the cometary perihelia. \footnote{Note that our angular distribution is different from the one given in \cite{matese11} because the direction of perihelion is opposite to that of aphelion.}
The two peaks in the longitude distribution suggest a great circle on the sky passing through $l=135^{\circ}$ and $l=315^{\circ}$ \citep{matese99,matese11}. 
We explain this anisotropy in Section \ref{sec:ADP}.

\section{Simulation of cometary orbits}\label{sec:simulation} 

We now build dynamical models of the Oort cloud comets and their perturbation via the Galactic tide and stellar encounters by simulating the passage of the solar system through the Galaxy.  We first introduce the Galactic potential, which yields a tidal gravitational force on the Sun and Oort Cloud comets. Then we give the initial conditions of the Oort cloud and the distribution of stellar encounters. Then we outline the numerical methods used to calculate the solar motion and the comet orbits.

\subsection{Galactic potential}\label{sec:potential} 

We adopt a Galactic potential with three components, namely an axisymmetric disk and a spherically symmetric halo and bulge
\begin{equation}
  \Phi_{\rm sym}=\Phi_b+\Phi_h+\Phi_d
  \label{eqn:Phi_sym}
\end{equation}
(this is same model as in \citet{feng13}).
The components are defined (in cylindrical coordinates) as
\begin{eqnarray}
  \Phi_{b,h}&=&-\frac{GM_{b,h}}{\sqrt{R^2+z^2+b_{b,h}^2}},\\
  \Phi_{d}&=&-\frac{GM_{d}}{\sqrt{R^2+(a_d+\sqrt{(z^2+b_d^2)})^2}},
  \label{eqn:Phi_component}
\end{eqnarray}
where $R$ is the disk-projected galactocentric radius of the Sun and $z$ is its vertical displacement above the midplane of the disk. $M$ is the mass of the component, $b$ and $a$ are scale lengths, and $G$ is the gravitational constant.  We adopt the values of these parameters from \citet{sanchez01}, which are listed in Table \ref{tab:model_par}.

\begin{table}
  \centering
    \caption{The parameters of the Galactic potential model for the symmetric component \citep{sanchez01}, the arm \citep{cox02,wainscoat92}, and the bar \citep{dehnen00}.}
    \label{tab:model_par}
    \begin{tabular}{@{}ll@{}}
      \hline
      component& parameter value \\\hline
      Bulge    & $M_b=1.3955 \times 10^{10}~M_\odot$ \\
      & $b_b =0.35$\,kpc \\
      Halo     & $M_h=6.9766\times 10^{11}~M_\odot$\\
      & $b_h=24.0$\,kpc\\
      Disk     & $M_d=7.9080\times 10^{10}~M_\odot$\\
      & $a_d=3.55$\,kpc\\
      & $b_d=0.25$\,kpc\\
      Arm    & $\zeta=15^\circ$\\
      & $R_{\rm min}=3.48$\,kpc\\
      & $\phi_{\rm min}=-20^\circ$\\
      & $\rho_0=2.5 \times 10^7 M_\odot {\rm kpc}^{-3}$\\
      & $r_0=8$\,kpc\\
      & $R_s=7$\,kpc\\
      & $H=0.18$\,kpc\\
      & $\Omega_s=20$\,km$s^{-1}$/kpc\\
      bar   & $R_b/R_{\rm CR}=0.8$\\
      & $\alpha=0.01$\\
      & $R_{\rm CR}=R_\odot(t=0~{\rm Myr})/2$\\
      & $\alpha=0.01$\\
      & $\Omega_b=60$\,km$s^{-1}$/kpc\\
      \hline
    \end{tabular}
\end{table}

In Section \ref{sec:sensitivity} we will add to this non-axisymmetric and
time-varying components due to spiral arms and the Galactic bar, to give the new potential
\begin{equation}
  \Phi_{\rm asym}=\Phi_{\rm sym}+\Phi_{\rm arm}+\Phi_{\rm bar} ~,
  \label{eqn:Phi_asym}
\end{equation}
where $\Phi_{\rm arm}$ is a potential of two logarithmic arms from
\citet{wainscoat92} with parameters given in \citet{feng13}, and
$\Phi_{\rm bar}$ is a quadrupole potential of rigid rotating bar from
\citet{dehnen00}. 
These components are used in the potential for the calculation of the solar orbit, but not the stellar encounter rate discussed in section~\ref{sec:encmod}.

The geometry of the arm is
\begin{equation}
  \phi_s(R) = \log(R/R_{\rm min})/\tan(\zeta)+\phi_{\rm min},
  \label{eqn:phi_spiral}
\end{equation}
where $\zeta$ is the pitch angle, $R_{\rm min}$ is the inner radius,
and $\phi_{\rm min}$ is the azimuth at that inner radius. 
A default pattern speed of $\Omega_p=20$ km~s$^{-1}$~kpc$^{-1}$ is adopted \citep{martos04,drimmel00}. The corresponding potential of this arm model is
\begin{eqnarray}
  \Phi_{arm} &=& -\frac{4 \pi G H}{K_1 D_1}\rho_0 e^{-\frac{R-r_0}{R_s}}\nonumber\\
            &&\times \cos(N[\phi-\phi_s(R,t)])\left[\rm{sech}\left(\frac{K_1z}{\beta_1}\right)\right]^{\beta_1}~,
  \label{eqn:Phi_arm}
\end{eqnarray}
where 
\begin{eqnarray*}
  K_1&=&\frac{N}{R \sin \zeta},\\
  \beta_1 &=& K_1 H (1+0.4 K_1 H),\\
  D_1 &=&\frac{1+K_1 H +0.3 (K_1 H)^2}{1+0.3 K_1 H},
\end{eqnarray*}
and $N$ is the number of spiral arms. The parameters in equation \ref{eqn:Phi_arm} are given in Table \ref{tab:model_par}.

The bar potential is a 2D quadrupole \citet{dehnen00}. Because the
Sun always lies outside of the bar, we adopt the potential
\begin{equation}
\Phi_{bar}=-A_b \cos[2(\phi-\Omega_b t-\phi_{\rm min})] \left[\left(\frac{R}{R_b}\right)^3 - 2\right]  \ \ \ R \geq R_b
\label{eqn:Phib_geqRb}
\end{equation}
where $R_b$ and $\Omega_b$ are the size and pattern speed of the bar
respectively and $\phi_{\rm min}$ is the bar angle. We assume that the spiral arms start from the ends of the major axis of the bar.
We only consider the barred state and ignore the evolution of the bar, so we
adopt a constant amplitude for the quadrupole potential, i.e.\ $A_b = A_f
$, in equation (3) of \citet{dehnen00}. $A_f$ is determined by the definition of the bar strength
\begin{equation}
\alpha\equiv 3~\frac{A_f}{v^2}\left(\frac{R_b}{R}\right)^3,
\label{eqn:bar_strength}
\end{equation}
where $R$ and $v$ are the current galactocentric distance of the Sun and
the corresponding local circular velocity. 
The fixed bar strength is given in Table \ref{tab:model_par}, from which we calculate $A_f$ and hence $A_b$.

\subsection{\label{sec:oort} Oort Cloud}

We generate Oort cloud comets using two different models, one from \citet{duncan87} (hereafter DQT) with the parameters defined in \citet{rickman08}, and another which we have reconstructed from the work of \cite{dones04b} (hereafter DLDW).

In the DQT model, initial semi-major axes ($a_0$) for comets are selected randomly from the interval $[3000,~10^5]$\,AU with a probability
density proportional to $a_0^{-1.5}$. The initial eccentricities ($e_0$) are
selected with a probability density proportional to $e_0$ \citep{hills81}, in such a way that the perihelia ($q_0$) are guaranteed to be larger than 32 AU. We generate the other orbital elements --- $\cos i_0$, $\omega_0$, $\Omega_0$ and $M_0$ --- from uniform distributions.
Because the density profile of comets is proportional to $r^{-3.5}$, where $r$ is the sun-comet distance, about 20\% of the comets lie in the classical Oort Cloud ($a>20\,000$\,AU).

In the DLDW model, the initial semi-major axes, eccentricities, and inclination angles are generated by Monte Carlo sampling from the relevant distributions shown in \cite{dones04b}. This produces semi-major axes in the range 3000 to 100\,000\,AU and ensures that the perihelia are larger than 32\,AU. Unlike the DQT model, there is a dependency of the cometary eccentricity and inclination on the semi-major axis, as can be see in Figures 1 and 2 of \cite{dones04c}. We generate comet positions and velocities relative to the invariant plane and then transform these into vectors relative to the Galactic plane.  In doing so we adopted values for the Galactic longitude and latitude of the north pole of the invariant plane of $98^{\circ}$ and $29^{\circ}$ respectively.

\begin{figure}
  \centering
  \hspace*{-2mm}\includegraphics[scale=0.8]{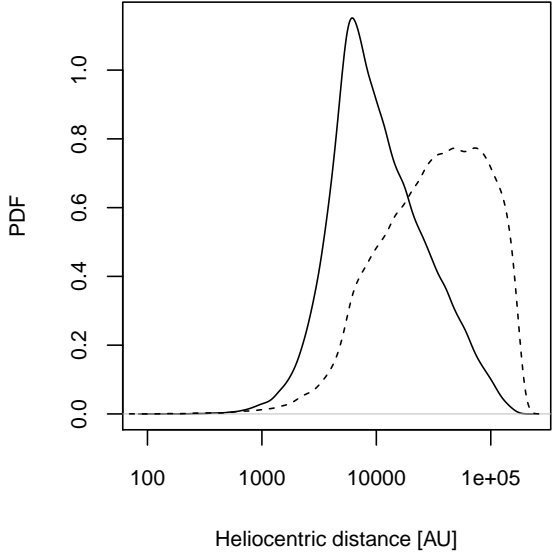}
  \caption{The normalized distributions of initial heliocentric distances of comets generated from the DQT model (solid line) and DLDW model (dashed line) with a sample size of $10^5$.}
  \label{fig:DQT_DLDW}
\end{figure}

The distributions of the cometary heliocentric distances for the DQT and DLDW models are given in Figure \ref{fig:DQT_DLDW}. We see that the DQT model produces more comets in the inner Oort cloud ($<$\,20\,000\,AU) and the DLDW model more in the outer Oort Cloud ($>$\,20\,000\,AU). Our distributions differ slightly from those in Figure 3 of \cite{dybczynski02} because our initial semi-major axes have different boundaries, and because our reconstruction of initial eccentricities and inclination angles is slightly different from the approach used in \cite{dybczynski02}. Many other Oort cloud initial conditions have been constructed numerically \citep{emelyanenko07,kaib11}.
Given the inherent uncertainty of the Oort cloud's true initial conditions, 
we carry out our work using two different Oort cloud models and investigate the sensitivity of our results to this (e.g.\ in section \ref{sec:ADP}).

\subsection{Stellar encounters}\label{sec:encmod} 

The geometry of encounters is complicated by the Sun's motion relative to the
local standard of rest (LSR). This solar apex motion could, by itself, produce
an anisotropic distribution in the directions of stellar encounters in the
heliocentric rest frame (HRF). Any anisotropy must be taken into account when trying to explain the observed anisotropic perihelia of the LPCs. Nonetheless, \citet{rickman08} simulated cometary orbits with an isotropic distribution of stellar encounters which is inconsistent with their method for initializing encounters. Here we use their method to generate encounters, but now initialize stellar encounters self-consistently to have a non-uniform angular distribution.

\subsubsection{Encounter scenario}\label{sec:scenario_enc}

\begin{figure}
  \centering
  \vspace{-10pt}
  \includegraphics[height=90mm,width=80mm]{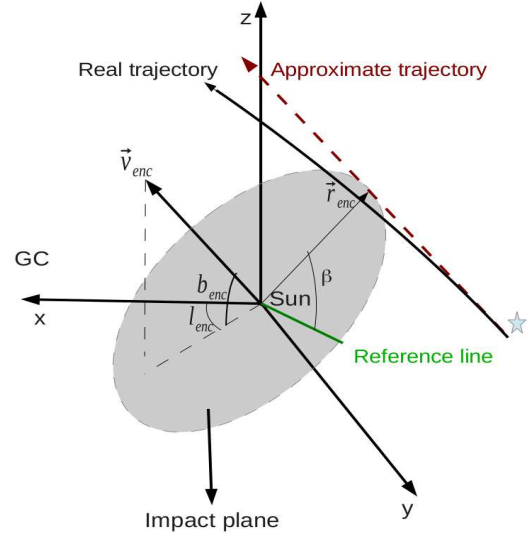}
  \vspace{-12pt}
  \caption{Schematic illustration in the heliocentric rest frame
    of stellar encounters. The circle is the impact plane which is
    defined by its normal, the encounter velocity $\vec{v}_{\rm enc}$. $\beta$
    is the angle in the impact plane measured from the reference axis
    to the stellar perihelion (i.e.\ the encounter). The vector in this plane
    from the Sun to the position of the encounter (i.e. the star’s perihelion)
    is defined as $r_{\rm enc}$. $b_{\rm enc}$ and $l_{\rm enc}$ are the
    Galactic latitude and longitude of $\vec{v}_{\rm enc}$,
    respectively. $(x,y,z)$ is the Galactic coordinate system. $\vec{r}_{\rm enc}$ is defined as the shortest distance from
    the Sun to the approximate trajectory which is a straight line in the
    direction of $\vec{v}_{\rm enc}$. The approximate trajectory of an
    encounter is used for the definition of encounter perihelion $\vec{r}_{\rm enc}$ while the real trajectory is integrated through simulations.} 
  \label{fig:impact_frame}
\end{figure}

The parameters of stellar encounters are generated using a Monte Carlo
sampling method, as follows. We distribute the encounters into different stellar
categories (corresponding to different types of stars) according to their
frequency, $F_i$, as listed in Table 8 of \citet{sanchez01}. In each stellar
category, the stellar mass $M_i$, Maxwellian velocity dispersion $\sigma_{\star i}$, and solar peculiar velocity $v_{\odot i}$, are given. The encounter scenario in the HRF is illustrated
in Figure \ref{fig:impact_frame}. The encounter perihelion $\vec{r}_{\rm enc}$ direction (which has Galactic coordinates $b_p$ and $l_p$) is by definition perpendicular to the encounter velocity $\vec{v}_{\rm
  enc}$. The angle $\beta$ is uniformly distributed in the interval of
$[0,2\pi]$. 

In this encounter scenario in the HRF, the trajectory of a stellar encounter is
determined by the encounter velocity $\vec{v}_{\rm enc}$, the encounter perihelion
$\vec{r}_{\rm enc}$, and the encounter time $t_{\rm enc}$. 
In the following paragraphs, we will first find the probability density
function (PDF) of encounters for each stellar category as a function of
$t_{\rm enc}$, $r_{\rm enc}$, and $v_{\rm enc}$, and then sample these parameters from this using the Monte
Carlo method introduced by \cite{rickman08} (hereafter R08). Then we will
sample $b_{\rm enc}$ and $l_{\rm enc}$ using a revised version of R08's method. Finally, $b_p$ and $l_p$ can be easily sampled because $\vec{r}_{\rm enc}$ is perpendicular to $\vec{v}_{\rm enc}$.

\subsubsection{Encounter probability}\label{sec:pdf_enc}

The probability for each category of stars is proportional to the number of stars passing through a ring with a width of $dr_{\rm enc}$ and centered on the Sun. The non-normalized PDF is therefore just
\begin{equation}
  P_u(t_{\rm enc}, r_{\rm enc}, v_{\rm enc}) \,=\, 4 \pi n_i v_{\rm enc} r_{\rm enc} \,\propto\, \rho(t_{\rm enc})v_{\rm enc}r_{\rm enc} ,
  \label{eqn:PDF_enc}
\end{equation}
where $n_i$ is the local stellar number density of the $i^{th}$ category of
stellar encounters, and $\rho(t_{\rm enc})$ is the local stellar mass density, which will change
as the Sun orbits the Galaxy.\footnote{We assume
  that the mass densities of different stellar categories have the same
  spatial distribution.} Thus the encounter probability is proportional to the
local mass density, the encounter velocity and the encounter perihelion.
We use a Monte Carlo method to sample $t_{\rm enc}$, $v_{\rm enc}$, and $r_{\rm enc}$ from this.

In different application cases, we sample the encounter time $t_{\rm enc}$
over different time spans according to equation \ref{eqn:PDF_enc}, where the
local mass density is calculated using Poisson's equation with the potentials
introduced in section \ref{sec:potential}.
Although we may simulate stellar encounters over a long time scale, we ignore
the change of the solar apex velocity and direction when simulating
the time-varying comet flux (in section \ref{sec:impact}) and the angular
distribution of current LPCs (in section \ref{sec:ADP}).
We select $r_{\rm enc}$ with a PDF proportional to $r_{\rm enc}$ with an upper
limit of $4\times 10^5$ AU. However, the sampling process of $v_{\rm enc}$ is
complicated by the solar apex motion and the stellar velocity in LSR, which we accommodate in the following way.

The encounter velocity in the HRF, $\vec{v}_{\rm enc}$, is the
difference between the velocity of the stellar encounter in the LSR, $\vec{v}_\star$, and the solar apex velocity relative to that type of star (category $i$)
in the LSR, $\vec{v}_{\odot i}$, i.e.\footnote{We
define a symbol without using the subscript $i$ when the symbol is derived
from a combination of symbols belonging and not belonging to certain stellar category.} 
\begin{equation}
  \vec{v}_{\rm enc}=\vec{v}_\star-\vec{v}_{\odot i}~.
  \label{eqn:vector_venc}
\end{equation}
We can consider the above formulae as a transformation of a stellar
velocity from the LSR to the HRF. The magnitude of this velocity in the HRF is
\begin{equation}
  v_{\rm enc}=[v^2_\star+v^2_{\odot i}-2v_{\odot i}v_\star \cos \delta]^{1/2}~,
  \label{eqn:venc}
\end{equation}
where $\delta$ is the angle between $\vec{v}_\star$ and $\vec{v}_{\odot i}$ in
the LSR.

To sample $v_{\rm enc}$, it is necessary to take into account both the encounter probability given in equation \ref{eqn:PDF_enc} and the distribution of $v_\star$. We generate $v_\star$ using
\begin{equation}
  v_\star = \sigma_{\star i} \left[\frac{1}{3}(\eta_u^2+\eta_v^2+\eta_w^2)\right]^{1/2},
  \label{eqn:vstar}
\end{equation}
where $\sigma_{\star i}$ is the stellar velocity dispersion in the $i^{\rm th}$ category, and $\eta_u$, $\eta_v$, $\eta_w$ are random variables, each
following a Gaussian distribution with zero mean and unit variance.

We then realize the PDF of encounters over $v_{\rm enc}$ (i.e.\ $P_u \propto v_{\rm enc}$) using R08's method as follows: (i) we randomly generate $\delta$ to be uniform in the interval $[0,2\pi]$; (ii) adopting $v_{\sun i}$ from table 1 in R08 and generating $v_\star$ from equation \ref{eqn:vstar}, we calculate $v_{\rm enc}$ using equation \ref{eqn:venc}; (iii) we define a large velocity $V_{\rm enc}=v_{\odot i}+3\sigma_{\star i}$ for the relevant star category and randomly draw a velocity $v_{\rm rand}$ from a uniform distribution over $[0, V_{\rm enc}]$. If $v_{\rm rand} < v_{\rm enc}$, we accept $v_{\rm enc}$ and the values of the generated variables $\delta,v_\star$. Otherwise, we reject it and repeat the process until $v_{\rm rand} < v_{\rm enc}$.

\begin{figure}
  \centering
  \includegraphics[scale=0.8]{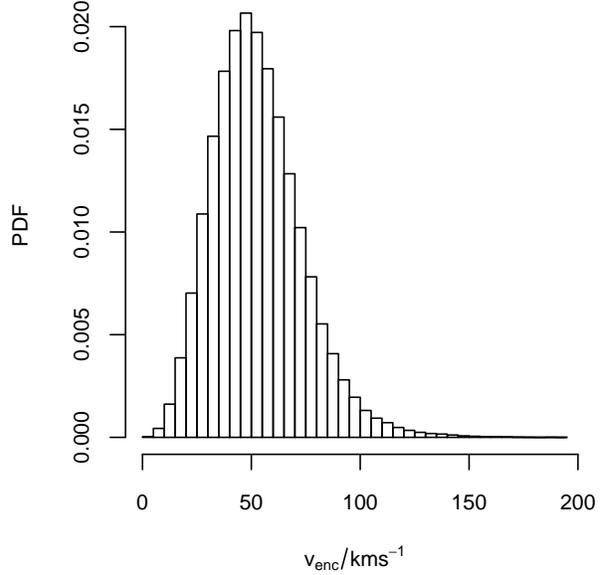}
  \caption{The histogram of the distribution of $v_{\rm enc}$ of all types of stars. 
The total number of encounters is 197\,906, which is the set of simulated encounters over the past 5\,Gyr.} 
  \label{fig:venc}
\end{figure}

We generate $10^5$ encounters in this way. Figure \ref{fig:venc} shows the resulting distribution of $v_{\rm enc}$. It follows a positively-constrained Gaussian-like distribution with mean velocity of 53\,km/s and a dispersion of 21\,km/s, which is consistent with the result in R08. In their modelling, R08 adopt a uniform distribution for $\sin b_{\rm enc}$, and $l_{\rm enc}$. This is not correct, however, because encounters are more common in the direction of the solar antapex where the encounter velocities are larger than those in other directions (equation \ref{eqn:PDF_enc}). We will show how to find the true distribution of $\sin
b_{\rm enc}$, $l_{\rm enc}$, $\sin b_p$ and $l_p$ as follows.

\subsubsection{Anisotropic perihelia of encounters}\label{anisotropic}

To complete the sampling process of encounters, we need to find a 5-variable
PDF, i.e. $P_u(t_{\rm enc}, r_{\rm enc}, v_{\rm enc}, b_{\rm enc}, l_{\rm
  enc})$. We have used R08's original Monte Carlo method to generate $t_{\rm
  enc}$, $r_{\rm enc}$ and $v_{\rm enc}$ according to equation
\ref{eqn:PDF_enc}. However, $b_{\rm enc}$ and $l_{\rm enc}$ are not generated
because R08 only use equation \ref{eqn:venc} to generate the magnitude of
$\vec{v}_{\rm enc}$ rather than the direction of $\vec{v}_{\rm enc}$. To
sample the directions of $\vec{v}_{\rm enc}$, we change the first and second
steps in R08's method introduced in section \ref{sec:pdf_enc} as follows: (i) we
randomly generate $\{b_\star, l_\star\}$ such that $\sin b_\star$ and
$l_\star$ are uniform in the interval of $[-1,1]$ and $[0,2\pi]$,
respectively; (ii) adopting $b_{\rm apex}=58.87^\circ$ and $l_{\rm apex}=17.72^\circ$ for the solar apex direction and generating $v_\star$ according to equation \ref{eqn:vstar}, we calculate $\vec{v}_{\rm enc}$ according to equation \ref{eqn:vector_venc}. 

\begin{figure}
  \centering
  \includegraphics[scale=0.45,angle=-90]{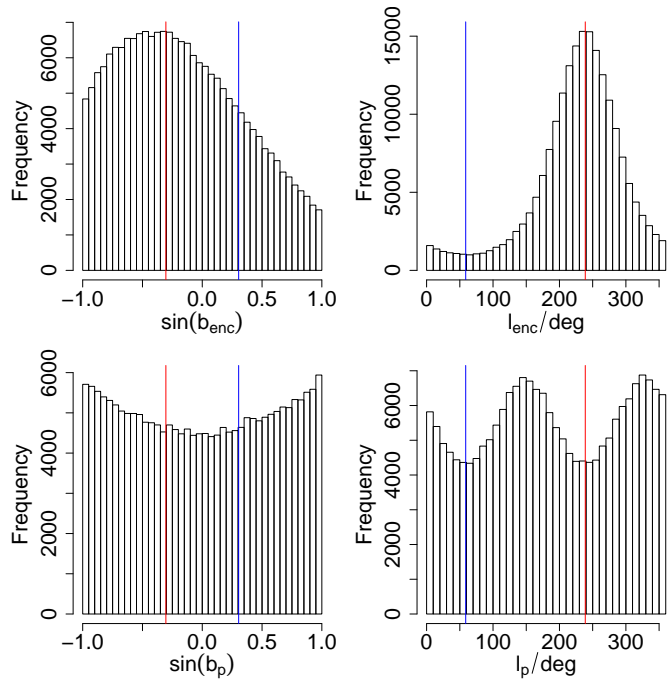}
  \caption{The upper panels show the distributions of the directions of the stellar encounter velocities in our simulations in Galactic coordinates as $\sin b_{\rm enc}$ (upper left)
    and $l_{\rm enc}$ (upper right). The lower panels show the distributions of
the directions of the corresponding perihelia as
    $\sin(b_p)$ (lower left) and $l_p$ (lower right). The blue and red lines
    denote the apex and antapex directions, respectively. The total number of encounters is 197\,906, which is the set of simulated encounters over the past 5 Gyr.} 
  \label{fig:venc_denc_bl}
\end{figure}

Selected in this way, $\sin b_\star$, $l_\star$, $\sin b_{\rm enc}$, and
$l_{\rm enc}$ all have non-uniform distributions. The Galactic latitude $b_p$ and
longitude $l_p$ of the encounter perihelia are also not uniform. 
Like R08, we draw 197\,906 encounters over the past 5 Gyr
from our distribution of encounters. The resulting histograms of $\sin b_{\rm enc}$, $l_{\rm enc}$, $\sin b_p$, and $l_p$ are shown in Figure \ref{fig:venc_denc_bl}. 
We see that the encounter velocity, $\vec{v}_{\rm enc}$, concentrates in the
antapex direction, while the encounter perihelion, $\vec{r}_{\rm enc}$,
concentrates in the plane perpendicular to apex-antapex direction. In
addition, the distribution of $l_p$ is flatter than that of $l_{\rm enc}$
because $\vec{r}_{\rm enc}$ concentrates on a plane rather than along a direction.

\begin{figure}
  \centering
  \includegraphics[scale=0.8,angle=-90]{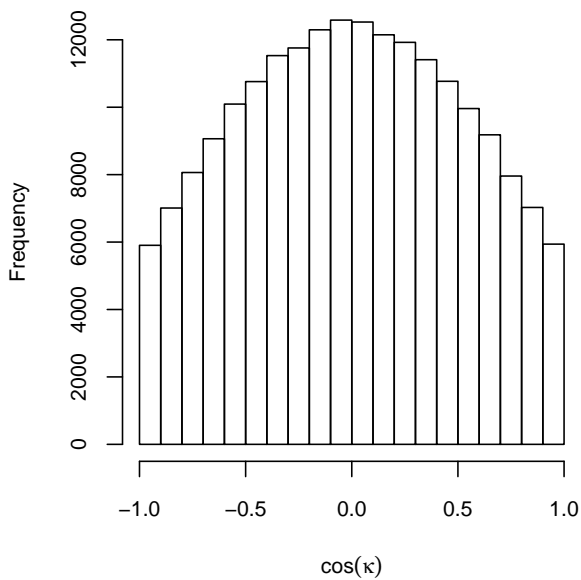}
  \caption{The distribution of the cosine of the angle between the encounter
    perihelion and the solar apex.}
  \label{fig:apex_kappa}
\end{figure}

In order to clarify the effect of the solar apex
motion, we define $\kappa$ as the angle between the encounter perihelion
$\vec{r}_{\rm enc}$ and the solar apex. If there were no solar apex motion, $\cos \kappa$
would be uniform. The effect of solar apex motion is shown in Figure
\ref{fig:apex_kappa}. The solar apex motion would result in the concentration
of encounter perihelia on the plane perpendicular to the apex direction. This
phenomenon is detected by \citet{sanchez01} using Hipparcos data, although the observational incompleteness biases the data. The non-uniform distribution over $\cos \kappa$ results in an anisotropy in the perihelia of LPCs, as we will demonstrate and explain in Section \ref{sec:ADP}.

\subsection{Methods of numerically simulating the comet orbits}\label{sec:method} 

\subsubsection{AMUSE}\label{sec:amuse}

Taking the above models and initial conditions, we
construct an integrator for the orbits of Oort cloud comets via a procedure
similar to that in \citet{wisdom91}, using the Bridge method \citep{fujii07}
in the AMUSE framework\footnote{http://www.amusecode.org} (a platform for coupling existing codes from
different domains; \citealp{pelupessy13,portegies13}). A direct integration of the cometary orbits is  computationally
expensive due to the high eccentricity orbits and the wide
range of timescales involved. We therefore split the dynamics of the comets into
Keplerian and interaction terms (following \citealp{wisdom91}). The Keplerian
part has an analytic solution for arbitrary time steps, while the interaction
terms of the Hamiltonian consist only of impulsive force kicks.
To achieve this we split the Hamiltonian for the system in the following way
\begin{equation}
  H = H_{\rm Kepler} + H_{\rm encounter} + H_{\rm tide}
  \label{eqn:H_split}
\end{equation}
where $H_{\rm Kepler}$, $H_{\rm encounter}$, and $H_{\rm tide}$ describe the
interaction of the comet with the dominant central object (the Sun), a passing
star, and the Galactic tide, respectively. Specifically, the
Keplerian cometary orbits can be integrated analytically according to $H_{\rm
  Kepler}$ while the interactions with the Galactic tide and stellar encounters are taken into account in terms of force kicks. For the time integration a 
second order leapfrog scheme is used, where the Keplerian evolution is 
interleaved with the evolution under the interaction terms. The forces for 
the latter are calculated using direct summation, in which the comet masses are neglected.
Meanwhile, the Sun moves around the
Galactic center under the forces from the Galactic tide and stellar encounters
calculated from $H_{\rm encounter}$ and $H_{\rm tide}$ in the leapfrog scheme.

We first initialize the orbital elements of the Sun and encountering
stars about the Galaxy, and the Oort cloud comets about the Sun.  We treat
the stellar encounters as a N-body system with a varying number of
particles, simulated using the Huayno code \citealt{pelupessy12}. The
interaction between comets and the Sun is simulated with a Keplerian code based
on \citet{bate71}.

At each time step in the orbital integration we calculate the gravitational force from the Galaxy and stellar encounters. The velocities of the comets are changed according to the Hamiltonian in equation \ref{eqn:H_split} at every half time step. Meanwhile, each comet moves in its Keplerian orbit at each time step. All variables are transformed into the HRF in order to take into account the influence of the solar motion and stellar encounters on the cometary orbits.

We use constant time steps in order to preserve the symplectic properties of the integration scheme in AMUSE (although we note that a symplectically corrected adaptive time step is used in some codes, such as SCATR \citep{kaib11b}). We use a time step of 0.1\,Myr for tide-only simulations because we find no difference in the injected flux when simulated using a smaller time step. 
The choice of time step size is a trade-off between computational speed and sample noise in the injected comet sample.  We use a time step of 0.01\,Myr in the encounter-only and in the combined (tide plus encounter) simulations when modelling the angular distribution of the LPCs' perihelia (section \ref{sec:ADP}).  (In section \ref{sec:sensitivity} we repeat some of these simulations with a shorter time step -- 0.001\,Myr -- to confirm that this time step is small enough.)  We use a time step of 0.001\,Myr in all other simulations.

In the following simulations we adopt the initial velocity of the Sun from \citet{schoenrich10} and the initial galactocentric radius from \citet{schoenrich12}. Other initial conditions and their uncertainties are the same as in \cite{feng13}. The circular velocity of the Sun (at $R=8.27$\,kpc), $v=225.06$\,km/s, is calculated based on the axisymmetric Galactic model in Section \ref{sec:potential}. These values are listed in Table \ref{tab:initial_condition}.
\begin{table*}
\caption{The current phase space coordinates of the Sun, represented as
  Gaussian distributions, and used as the initial conditions in our orbital
  model \citep{schoenrich10,schoenrich12,majaess09,dehnen98b}.}
\centering
\begin{tabular}{l*5{c}r}
\hline
\hline
     &$R$/kpc&$V_R$/kpc~Myr$^{-1}$&$\phi$/rad&$\dot\phi$/rad~Myr$^{-1}$&z/kpc&$V_z$
/kpc~Myr$^{-1}$\\
\hline
mean &8.27  &-0.01135           & 0        & 0.029&0.026&-0.0074\\
standard deviation &0.5& 0.00036&0         &0.003 &0.003&0.00038\\
\hline
\end{tabular}
\label{tab:initial_condition}
\end{table*}

\subsubsection{Numerical accuracy of the AMUSE-based method}\label{sec:accuracy}

\begin{figure}
\centering
\includegraphics[scale=0.4,angle=-90]{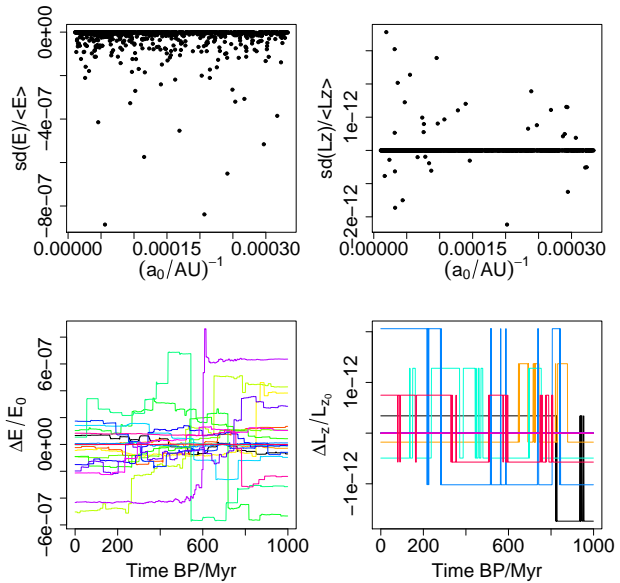}
\caption{Assessment of the numerical accuracy of the AMUSE-based method through monitoring the conservation of energy $E$ and angular momentum $L_z$ for 1000 comets generated from the DLDW Oort cloud model. Upper panels: For each of the 1000 comets, the standard deviation (over its orbit) of $E$ (left) and $L_z$ (right) relative to the average value over the orbit, plotted as a function of the initial energy (which is proportional to $1/a_0$). Lower panels: the fractional change over the orbit of $E$ and $L_z$ for the 20 comets (represented by different colours) with the highest numerical errors. } 
\label{fig:E_Lz_test}
\end{figure}

To test the numerical accuracy of the AMUSE-based method, we generated 1000 comets from the DLDW model and monitored the conservation of orbital energy and angular momentum. As the perturbation from the Galactic potential and stellar encounters used in our work would violate conservation of the third component of angular momentum ($L_z$), we use a simplified Galactic potential for this test, namely a massive and infinite sheet with
  \begin{equation}
    \Phi_{\rm sheet}=2\pi G\sigma |z|,
    \label{eqn:slab_disk}
  \end{equation}
where $G$ is the gravitational constant, $\sigma = 5.0\times 10^6~M_{\sun}/$\,kpc$^2$ is the surface density of the massive sheet and $z$ is the vertical displacement from the sheet. 
Because this potential imposes no tidal force on comets if the Sun does not cross the disk, it enables us to test the accuracy of the bridge method in AMUSE by using the conservation of cometary orbital energy and the angular momentum perpendicular to the sheet. To guarantee that the Sun does not cross the plane during the 1\,Gyr orbital integration (i.e.\ the oscillation period is more than 2\,Gyr), we adopt the following initial conditions of the Sun: $R=0$\,kpc, $\phi=0$, $z=0.001$\,kpc, $V_R=0$\,kpc/Myr, $\dot\phi=0$\,rad/Myr, $V_z=0.0715$\,kpc/Myr.
Integrating the cometary orbits over 1\,Gyr with a constant time step of 0.1\,Myr, we calculate the fractional change of the comets' orbital energies $E$ and the vertical component of their angular momenta $L_z$ during the motion (Figure \ref{fig:E_Lz_test}). Both quantities are conserved to a high tolerance, with fractional changes of less than $10^{-6}$ for $L_z$ and less than
$10^{-12}$ for $E$. The numerical errors are independent of the comet's energy (which is inversely proportional to the semi-major axis). Compared to the magnitude of the perturbations which inject comets from the Oort cloud into the observable zone, these numerical errors can be ignored during a 1\,Gyr and even a 5\,Gyr integration.

\subsubsection{Comparison of the AMUSE-based method with other methods}\label{sec:initial_condition}

Our numerical method calculates perturbations from stellar encounters
and the Galactic tide using dynamical equations directly, instead of
employing an impulse approximation (e.g.\ CIA, DIA, or SIA
\citet{rickman05}) or the Averaged Hamiltonian Method
(AHM)\citep{fouchard04}.  In the latter the Hamiltonian of the
cometary motion is averaged over one orbital period. This can
significantly reduce the calculation time, but is potentially less
accurate. A more explicit method is to integrate the Newtonian
equations of motion directly, e.g.\ via the Cartesian Method (CM) of
\citep{fouchard04}, but this is more time consuming.

\begin{figure}
\centering
\includegraphics[scale=0.8]{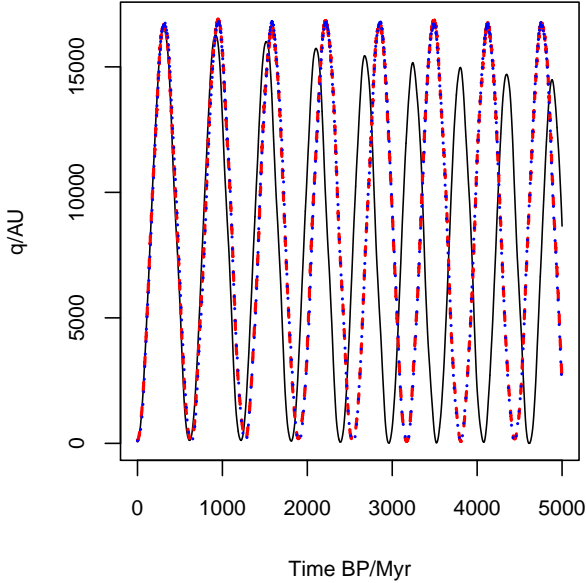}
\caption{The variation of the perihelion of one comet calculated with
  three different integration methods: AHM (black solid), CM (red dashed), and
  AMUSE-based method (blue dotted). } 
\label{fig:amuse_AHM}
\end{figure}

To illustrate the accuracy of the AHM, CM, and AMUSE-based methods in simulating high eccentricity orbits, we integrate the orbit of one comet using all methods. The test comet has a semi-major axis of $a=25\,000$\,AU and an eccentricity of $e=0.996$ (as used in \citet{fouchard04}). Adopting the following initial conditions of the Sun -- $R=8.0$\,kpc, $\phi=0$, $z=0.026$\,kpc, $V_R=-0.01$\,kpc/Myr, $\dot\phi=0.0275$\,rad/Myr, $V_z=0.00717$\,kpc/Myr -- and using the same tide model as described above, the solar orbit under the perturbation from the Galactic tide is integrated over the past 5\,Gyr.  Figure \ref{fig:amuse_AHM} shows that the evolutions of the cometary perihelia calculated using the CM and AMUSE-based methods are very similar, whereas AHM shows an evolution which diverges from these. As CM is the most accurate method, this shows that the
AHM cannot be used to accurately calculate the time-varying, because
it holds the perturbing forces constant during each orbit.
Because the AMUSE-based method computes a large sample of comets more
efficiently than CM does, we have adopted the AMUSE-based method in our work.

\subsubsection{Calculation of the injected comet flux}\label{sec:inject}

A comet which comes too close to the perturbing effects of the giant planets in the solar system will generally have its orbit altered such that it is injected into a much shorter periodic orbit or is ejected from the solar system on an unbound orbit. We regard a comet as having been injected into the inner solar system in this way when it enters into the ``loss cone'' \citep{wiegert99}, i.e.\ that region with a heliocentric radius of 15\,AU or less (the same definition as in \citet{dybczynski05} and R08). These are the comets which can then, following further perturbations from the planets, hit the Earth. If injected comets enter an observable zone within $<5$\,AU then they may be observed as a LPC. Comets which are injected into the loss cone or which are ejected from the solar system (i.e.\ achieve heliocentric distances larger than $4\times 10^5$\,AU) are removed from the simulation. 

The observable comets are only a subset of the injected comets because some injected comets can be ejected again by Saturn and Jupiter. But assuming that this is independent of the orbital elements over long time scales, we assume that the flux of injected comets is proportional to the flux of LPCs. 
Inner Oort cloud comets, in particular comets with $a < 3000$\,AU, may be injected into the loss cone  ($q < 15$\,AU) but not enter the observable zone ($q < 5$\,AU) \citep{kaib09}. 
In our simulations we will examine the properties of comets injected into both types of target zone, and we will refer to such injected comets as LPCs.
Once we have identified the injected comets, we calculate the Galactic latitudes $b_c$ and longitudes $l_c$ of their perihelia. Because the orbital elements of the class 1A LPCs are recorded during their first passage into inner solar system, we can reasonably assume that the direction of the LPC perihelion is unchanged after entering the ``loss cone''. In Section \ref{sec:impact} and \ref{sec:ADP}, we will model the terrestrial cratering time series and the anisotropic perihelion of LPCs based on the injected comet flux. Specifically, in Section \ref{sec:impact}, we will show how we convert the simulations of the perturbations of the cometary orbits into a model for the time variation of the cometary flux entering the inner solar system. 

\section{Bayesian inference method}\label{sec:bayesian} 

We summarize here our Bayesian method for quantifying how well a
time series model can describe a set of cratering data (or indeed any
other series of discrete time measurements with uncertainties).
A full description of the method and its application to the cratering
data for various non-dynamical
models can be found in \cite{bailer-jones11,2011MNRAS.418.2111B}. 

\subsection{Evidence}\label{sec:evidence} 

If we define $D$ as the time series of craters and $M$ as some model
for these data, then the evidence of the model is defined as 
\begin{equation}
  P(D|M) = \int_{\theta} P(D|\theta,M)P(\theta|M) d\theta,
  \label{eqn:evidence}
\end{equation}
where $\theta$ is the parameters of the model, and $P(D|\theta,M)$ and $P(\theta|M)$
are the likelihood of the data and the prior distribution over the
parameters, respectively.  The evidence is therefore the prior-weighted
average of the likelihood over the parameters. It gives the overall
ability of the model to fit the data, rather than the power of any
individual set of parameters. As is well known in statistics, and
further described in \cite{bailer-jones11}, this is the appropriate
metric to use in order to compare models of different flexibility or complexity.

If $t_j$ is the {\em true} (unknown) time of the impact of crater $j^{th}$, and 
$\tau_j$ is the {\em measured} time with corresponding uncertainty
$\sigma_j$, then an appropriate error model for this measurement is
\begin{equation}
  P(\tau_j|\sigma_j,t_j)=\frac{1}{\sqrt{2\pi}\sigma_j}\exp[-(\tau_j-t_j)^2/{2\sigma_j^2}]
  \ .
  \label{eqn:measurement}
\end{equation}
The likelihood for one crater measurement can then be calculated by
integrating over the unknown time
\begin{eqnarray}
  P(\tau_j|\sigma_j,\theta,M) &=&
  \int_{t_j}P(\tau_j|\sigma_j,t_j,\theta,M)P(t_j|\sigma_j,\theta,M)dt_j\nonumber\\
  &=&\int_{t_j} P(\tau_j|\sigma_j,t_j)P(t_j|\theta,M)dt_j \ .
  \label{eqn:event_like}
\end{eqnarray}
The second term in the second equation describes the time series model: it
predicts the probability that an event will occur at time $t_j$ given the
parameters for that model.
The likelihood for the whole time series, $D=\{\tau_j$\}, is the
product of the individual likelihoods (assuming they are measured
independently), in which case
\begin{equation}
  P(D|\theta,M)=\prod\limits_j P(\tau_j|\sigma_j,\theta,M) \ .
  \label{eqn:likelihood}
\end{equation}
We use this in equation \ref{eqn:evidence} to calculate the evidence
for model $M$ give the set of cratering dates.
The absolute scale of the evidence is unimportant: we are only
interested in ratios of the evidence for any pair of models, known as
the {\em Bayes factor}. As a rule of thumb, if the Bayes factor is
larger than 10, then the model represented in the numerator of the
ratio is significantly favoured by the data over the other model (see
\citet{kass95} for further discussion of the interpretation).

\subsection{Time series models}\label{sec:tsmodel}

The time series model, $M$, is a model which predicts the variation
of the impact probability with time (the normalized cratering
rate), i.e.\ the term $P(\tau_j|\sigma_j,\theta,M)$ in equation
\ref{eqn:likelihood}.  The models we use in this work, along with their
parameters, $\theta$, are defined in Table \ref{tab:tsmodels}, and described below 
  \begin{description}
  \item[] {\em Uniform.} Constant impact probability over the range of the data. As any probability distribution must be normalized over this range, this model has no parameters.
  \item[] {\em RandProb, RandBkgProb.} Both models comprise $N$ impact events at random times, with each event modelled as a Gaussian.
$N$ times are drawn at random from a uniform time distribution extending over the range of the data. 
A Gaussian is placed at each of these with a common standard deviation (equal to the average
of the real crater age uncertainties). We then sum the Gaussians, add a constant background, $B$, and normalize.
This is the RandBkgProb (``random with background'') model. RandProb is the special case for $B=0$.
We calculate the evidence by averaging over a large number of realizations of the model (i.e.\ times of the events), and, for RandBkgProb, over $B$. For example, when we later model the basic150 time series, we fix $N=32$ and range $B$ from 0 to $\infty$ (see Table \ref{tab:prior}).
  \item[] {\em SinProb, SinBkgProb.} Periodic model of angular frequency $\omega$ and phase $\phi_0$ (model SinProb). There is no amplitude parameter because the model is normalized over the time span of the data. Adding a background $B$ to this simulates a periodic variation on top of a constant impact rate (model SinBkgProb).
  \item[] {\em SigProb.} A monotonically increasing or decreasing nonlinear trend in the impact PDF using a sigmoidal function, characterized by the steepness of the slope, $\lambda$, and the center of the slope, $t_0$. In the limit that $\lambda$ becomes zero, the model becomes a step function at $t_0$, and in the limit of very large $\lambda$ it becomes the Uniform model. We restrict $\lambda <0$ in our model comparison because the decreasing trend in cratering rate towards the past seems obvious in the time series (see Figure \ref{fig:set3lim}; see also \cite{bailer-jones11}). However, we do include the increasing trend in our sensitivity test in Section \ref{sec:sensitivity}.
  \item[] {\em SinSigProb.} Combination of SinProb and SigProb.
  \item[] {\em TideProb, EncProb, EncTideProb.} Models arising from the dynamical simulation of cometary orbits perturbed by either stellar encounters (EncProb) or the Galactic tide (TideProb) or both (EncTideProb). We describe the modelling approach which produces these distributions in detail in Section \ref{sec:impact}.
  \item[] {\em EncSigProb, TideSigProb, EncTideSigProb.} Combination of EncProb, TideProb, EncTideProb (respectively) with SigProb.
  \end{description}

Some of these models -- those in the first five lines in Table \ref{tab:tsmodels} -- are simple analytic models. The others are models based on dynamical simulations of cometary orbits, which we therefore call dynamical models.
In the next section we will explain how we get from a simulation of the perturbation of the cometary orbits to a prediction of the cratering rate. Table \ref{tab:tsmodels} also lists the parameters of the models, i.e.\ those parameters which we average over in order to calculate the evidence. The prior distributions for these parameters are listed in Table \ref{tab:prior}.
\begin{table*}
\centering
\caption{The mathematical form of the time series models and their
  corresponding parameters. Time $t$ increases into the past and
  $P_u(t|\theta, M)$ is the unnormalized cratering rate (probability density)
  predicted by the model. In the dynamical models (EncProb, TideProb,
  EncTideProb, EncSigProb, TideSigProb, and EncTideSigProb), $\vec{r}_\odot(t=0Myr)$ and $\vec{v}_\odot(t=0Myr)$ are Sun's current position and velocity relative to
  the Galactic center. Note that the components in the compound models
  are normalized before being combined. The quantities $\gamma_{\rm bin}(t)$, $G_3(t)$, and $\xi$ are defined in Section
  \ref{sec:impact}. 
  $\eta$ is a parameter which
  describes the relative contribution of the two combined models.}
\begin{tabular}{l c r}
\hline
\hline
model name&$P_u(t|\theta,M)$&parameters, $\theta$\\
\hline
Uniform&1&none\\
RandProb/RandBkgProb&$\sum_{n=1}^{N}\mathcal{N}(t; \mu_n,\sigma)$+$B$&$\sigma$, $B$,$N$\\
SinProb/SinBkgProb&$1/2\{\cos[\omega t+\phi_0]+1\}$+$B$&$\omega$, $\beta$, $B$\\
SigProb &$[1+e^{(t-t_0)/\lambda}]^{-1}$&$\lambda$, $t_0$\\
SinSigProb &SinProb+SigProb&$T$, $\beta$, $B$,$\lambda$, $t_0$\\
EncProb &$\gamma_{\rm bin}(t)$&$\vec{r}_\odot(t=0)$, $\vec{v}_\odot(t=0)$\\
TideProb & $G_3(t)$ & $\vec{r}_\odot(t=0)$, $\vec{v}_\odot(t=0)$\\
EncTideProb &$[\gamma_{\rm bin}(t)+\xi G_3(t)]/(1+\xi)$&$\xi$, $\vec{r}_\odot(t=0)$, $\vec{v}_\odot(t=0)$\\
EncSigProb &EncProb + $\eta$ SigProb & $\eta$, $\lambda$, $t_0$, $\vec{r}_\odot(t=0)$, $\vec{v}_\odot(t=0)$\\ 
TideSigProb &TideProb + $\eta$ SigProb & $\eta$, $\lambda$, $t_0$, $\vec{r}_\odot(t=0)$, $\vec{v}_\odot(t=0)$\\ 
EncTideSigProb &EncTideProb + $\eta$ SigProb&$\xi$, $\eta$, $\lambda$, $t_0$, $\vec{r}_\odot(t=0)$, $\vec{v}_\odot(t=0)$\\
\hline
\end{tabular}
\label{tab:tsmodels}
\end{table*}

\begin{table*}
\centering
\caption{The prior distribution and range of parameters for the various time series
  models.
  For the non-dynamical models (i.e.\ all except the
  last five lines), a uniform prior for all
  the parameters is adopted which is constant inside the range shown
  and zero outside. 
  $N_{\rm ts}$ and $\tau_{\rm max}$ are the number of events and the earliest
  time of occurrence of the craters. $\bar{\sigma_i}$ is the averaged
  age uncertainties of the craters.
  The prior PDFs over the parameters of the dynamical
  models (the last five lines) are Gaussian, with means and standard
  deviations set by the
  initial conditions as listed in Table \ref{tab:initial_condition}. 
}
\begin{tabular}{l c}
\hline
\hline
model name & details of the prior over the parameters \\
\hline
Uniform& no parameters\\
RandProb &$\sigma=\bar{\sigma_i}$, $N=N_{\rm ts}$, $B=0$\\
RandBkgProb&$\sigma=\bar{\sigma_i}$, $N=N_{\rm ts}$, $B=\frac{1}{\sqrt{2\pi}\sigma}\frac{b}{(1-b)}$ with $b\in[0,1]$\\
SinProb&$2\pi/100 <\omega<2\pi/10$, $0<\phi_0<2\pi$,$B=0$\\
SinBkgProb&$2\pi/100 <\omega<2\pi/10$, $0<\phi_0<2\pi$, $B=\frac{b}{(1-b)}$ with $b\in[0,1]$\\
SigProb &$-100<\lambda<0$, $0<t_0<0.8\tau_{\rm max}$\\
SinSigProb &Priors from both SinProb and SigProb\\
EncProb & Initial conditions listed in Table \ref{tab:initial_condition} \\
TideProb & Initial conditions listed in Table \ref{tab:initial_condition}\\ 
EncTideProb & $\xi=1$, Initial conditions listed in Table \ref{tab:initial_condition}\\
EncSigProb& $0<\eta<4$, $-100<\lambda<0$, $0<t_0<0.8\tau_{\rm max}$, initial conditions listed in Table \ref{tab:initial_condition}\\
TideSigProb& $0<\eta<4$, $-100<\lambda<0$, $0<t_0<0.8\tau_{\rm max}$, initial conditions listed in Table \ref{tab:initial_condition}\\ 
EncTideSigProb&$\xi=1$, $0<\eta<4$, $-100<\lambda<0$, $0<t_0<0.8\tau_{\rm max}$, initial conditions listed in Table \ref{tab:initial_condition}\\ 
\hline
\end{tabular}
\label{tab:prior}
\end{table*}

\section{Modelling the history of the cometary impact rate}\label{sec:impact} 

The terrestrial impact rate consists of two parts: the asteroid impact
rate and the comet impact rate. We are specifically interested in 
only the latter in the present work. The background asteroid impact rate is
proportional to the number of asteroids in the asteroid belt, which is
depleted by the impact of asteroids on planets and their
satellites. Over a long time scale (longer than 100 Myr), the
background impact rate of asteroids would therefore decrease towards the present.
But we could also see variations in this
due to the disruption of large asteroids into an asteroid family, which would produce phases of enhanced impacting \citep{bottke07}.  
In addition to the actual impact rate, the geological record of all impact craters (comet or asteroid) is contaminated by a selection bias: The older a crater is, the more likely it is to have been eroded and so the less likely it is to be discovered. This preservation bias would lead to an apparent increase in the impact rate towards to the present.  We model the combined contribution of these two components (variable asteroid impact rate and the preservation bias)  to the measured impact rate using a sigmoidal function, which produces a smoothly varying trend with time (model SigProb in Table \ref{tab:tsmodels}). As with the other models, this model has parameters which we average over when computing the model evidence.

The cometary impact rate is determined by the
gravitational perturbations of the Oort cloud due to the Galactic tide
and stellar encounters. Both are modulated by the solar
motion around the Galactic center. Some studies suggest that their
combined effect injects more comets into the inner solar system
 than does each acting alone \citep{heisler87, rickman08}. This so-called synergy
effect is difficult to model, however, and will be ignored in our
statistical approach.

We simulate the effects of the tide and encounters separately (section~\ref{sec:simulation}). The resulting
cometary flux from these is described by
the models TideProb and EncProb respectively. The cometary
flux when both processes operate, the model EncTideProb, is the sum of the
fluxes from each (each being normalized prior to combination). 
To include the contributions from the asteroid
impacts and the crater preservation bias we can add to this the
SigProb model mentioned above. This gives the model EncTideSigProb.
The parameters of all these models and their prior ranges are defined
in Tables \ref{tab:tsmodels} and \ref{tab:prior}.

\subsection{Tide-induced cometary flux}\label{sec:tideflux}

The time variation as the Sun orbits the Galaxy of the tide-induced
cometary flux entering the loss cone is calculated using AMUSE-based method (section \ref{sec:method}).
We define $f_c$ as the relative injected comet flux in a time bin
with width $\Delta t$
\begin{equation}
  f_c=\frac{N_{\rm inj}}{N_{\rm tot}\Delta t},
  \label{eqn:f_tide}
\end{equation}
where $N_{\rm inj}$ is the number of injected comets in this bin and
$N_{\rm tot}$ is the total number of the comets. 

We could use $f_c$ directly as the model prediction of the
comet impact cratering rate, $P_u(t|\theta,M)$, for the model TideProb
(section~\ref{sec:tsmodel}) for that particular set of model parameters.
However, as the calculation of the cometary orbits is rather time-consuming,
we instead use a proxy for $f_c$, i.e. the vertical tidal force. 

The tidal force per unit mass experienced by a comet in the Oort Cloud is 
\begin{equation}
  \mathbf{F}=-\frac{G M_\odot \,\mathbf{\hat r}}{r^2} - G_1 x \,\mathbf{\hat x}  -G_2 y \, \mathbf{\hat y} - G_3 z \mathbf{\hat z}
  \label{eqn:tidal_force}
\end{equation}
where $\mathbf{r}$ is the Sun-comet vector of length $r$, $M_\odot$ is the
solar mass, and $G$ is the gravitational constant.\footnote{We don't use this
  equation in simulating cometary orbits in the AMUSE framework.} The three tidal coefficients, $G_1$, $G_2$, and $G_3$ are defined as
\begin{equation}
  \begin{array}{l}
  \displaystyle G_1=-(A-B)(3A+B)\\
  \displaystyle G_2=(A-B)^2\\
  \displaystyle G_3=4\pi G \rho(R,z)-2(B^2 -A^2)
  \end{array}
  \label{eqn:G123}
\end{equation}
where $A$ and $B$ are the two Oort constants, and $\rho(R,z)$ is the local mass
density
which can also be denoted as $\rho(t)$ in the case of using $G_3(t)$ to build models. Because the two components $G_1$ and $G_2$ in the Galactic $(x,y)$ plane are about ten times smaller than the vertical component ($G_3$), it is the vertical tidal force that dominates the perturbation of the Oort Cloud.

\begin{figure}
  \centering
  \includegraphics[scale=0.8]{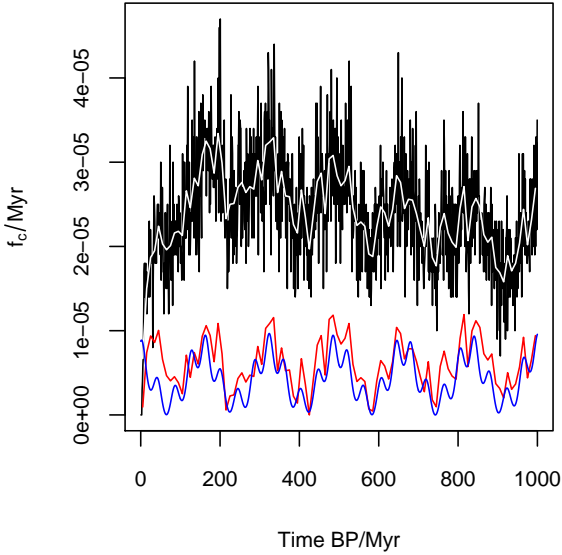}
  \caption{Comparison between the tide-induced injected comet flux ($f_c$) and the vertical Galactic tide ($G_3$). The injected comet flux is shown as a histogram with two different bins sizes: 1\,Myr (black line) and 10\,Myr (white line). The red line is the detrended comet flux with a time bin of 10\,Myr. The blue line shows the variation of $G_3$ (scaled, as it has a different unit to $f_c$). }
\label{fig:Fc_G3}
\end{figure}

To find a relationship between $f_c$ and $G_3$, we simulate the orbits of one million comets generated from the DQT model back to 1\,Gyr in the past under the perturbation of the Galactic tide (stellar encounters are excluded).  We use here the loss cone as the target zone when identifying the injected comets (LPCs).  The two quantities are compared in Figure \ref{fig:Fc_G3}. We see that the detrended comet flux (red line) agrees rather well with $G_3$ (blue line) over the past 1\,Gyr, albeit with an imperfect detrending over the first 100\,Myr.  We made a similar comparison for the DLDW model and also find a very close linear relation. Comparing $G_3$ with the flux of the comets injected into the observable zone (i.e. $q < 5$\,AU) for both the DLDW and DQT models, we find that the result is consistent with what we have found for the loss cone. This confirms the relationship between the tide-induced comet flux and the vertical tidal force, which was also demonstrated by \cite{gardner11} (their Figure 9) with a different approach. We are therefore justified in using $G_3$ as a proxy for the tide-induced comet flux when we build models of cometary impact rate to compare to the crater time series.

\subsection{Encounter-induced cometary flux}\label{sec:encounterflux}

We define the encounter-induced flux entering the loss cone in the same way as $f_c$ in equation \ref{eqn:f_tide}. We now investigate whether we can introduce a proxy for this too. We postulate the use of the quantity \begin{equation}
  \gamma=\frac{M_{\rm enc}}{v_{\rm enc}r_{\rm enc}}
  \label{eqn:gamma}
\end{equation}
which is proportional to the change in velocity of the Sun (or equivalently to the mean change in velocity of the comets) as induced by an encounter according to the classical impulse approximation \citep{oort50,rickman76}. This proxy has also been used in previous studies to approximate the LPC flux injected by stellar encounters (e.g.\ \citet{kaib09,fouchard11}).

\begin{figure}
  \centering
  \includegraphics[scale=0.8]{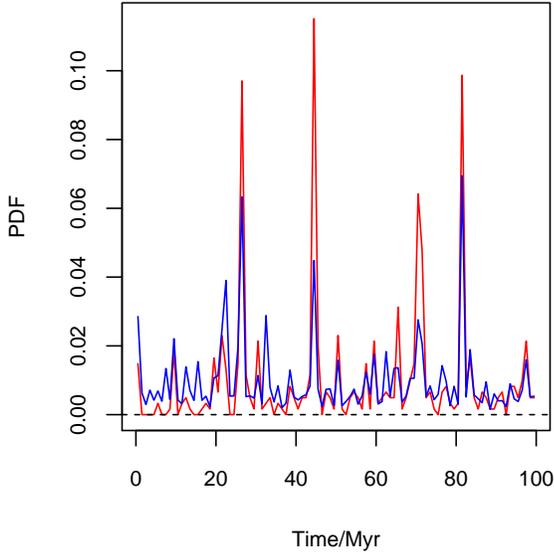}
  \caption{The time-varying probability density of the encounter-induced injected comet flux $f_c$ (red line) and the prediction of proxy $\gamma_{\rm bin}$ (blue line), binned with a time bin of 1\,Myr. }
  \label{fig:gamma_fc_pdf}
\end{figure}

The injected flux is dominated by those encounters which can signifcantly change the velocity and thus the perihelion of the comets
%the most
\citep{hills81,heisler87,fouchard11}. Considering the important role of these encounters and the long time scale between them (about 100 Myr according to \citealt{heisler87}), we divide the whole time span of simulated stellar encounters into several time bins and use the (normalized) maximum value of $\gamma$ in each bin to approximate such comet showers. We define this binned proxy as $\gamma_{\rm bin}$, and normalize it over the whole time scale. In Figure \ref{fig:gamma_fc_pdf}, we compare this proxy to the normalized encounter-induced flux which is simulated with a time step of 0.001\,Myr using a sample of $10^5$ comets generated from the DLDW model over 100\,Myr. We find that the main comet showers can be properly predicted by $\gamma_{\rm bin}$, although it may miss small comet showers and predict some non-existent small showers. 

\begin{figure}
  \centering
  \includegraphics[scale=0.8]{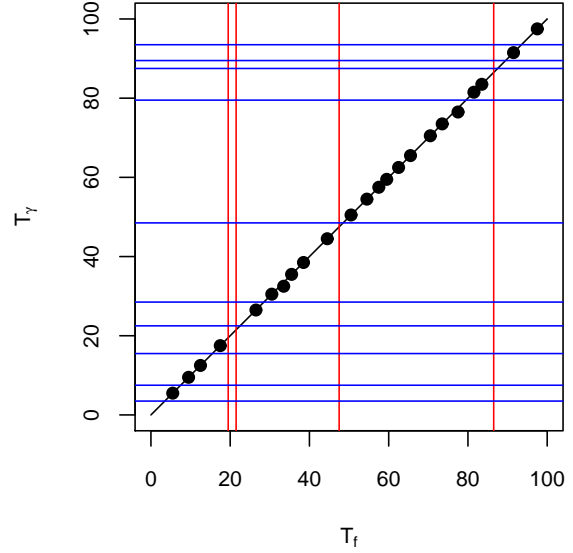}
  \caption{Assessment of the comet shower prediction ability of the proxy $\gamma$. The black points show peaks which are correctly reproduced, by plotting their time of occurrence in the proxy, $T_{\gamma}$, against their true time of occurrence, $T_f$, in $f_c$. Peaks missed by the proxy are shown as vertical red lines and false peaks in the proxy are shown as horizontal blue lines.}
  \label{fig:gamma_fc_peaks}
\end{figure}

To assess the reliability of the shower prediction of the proxy, we evaluate the fraction of
peaks in $f_c$ which are correctly identified by $\gamma_{\rm bin}$, and the fraction of peaks in
$\gamma_{\rm bin}$ which have a corresponding true peak in $f_c$. For the former case, a peak in $f_c$ is counted as correctly predicted by the proxy when it occurs in the same time bin as a peak in $\gamma_{\rm bin}$, or when the $f_c$ peak is one bin earlier (because the shower can occur up to 1\,Myr after the closest approach of the encounter). We find that 23 out of 27 (0.85) flux peaks are correctly predicted by the proxy, 
while 23 out of 33 (0.70) peaks in $\gamma_{\rm bin}$ have corresponding peaks in $f_c$ (Figure \ref{fig:gamma_fc_peaks}).
This simple counting ignores the intensity of the comet showers.
To remedy this use the amplitude of each $\gamma_{\rm bin}$ peak as a weight, 
and count the weighted fractions. We find these to be 0.92 and 0.84 respectively. These results suggests that $\gamma_{\rm bin}$ is a reasonably good proxy for statistical purposes.
Hence we use $\gamma_{\rm bin}$ as the measure of $P_u(t|\theta,M)$ for the model EncProb.
The linear relationship between $\rho(t)$ and $G_3(t)$ (equations \ref{eqn:PDF_enc} and \ref{eqn:G123}) indicates that the averaged EncProb model over sequences of $\gamma_{\rm bin}$ is equivalent to the corresponding TideProb model for one solar orbit. We will see in section \ref{sec:comparison} whether there is any significant difference between the evidences for these two models.

\subsection{Combined tide--encounter cometary flux}

\begin{figure}
  \centering
  \includegraphics[scale=0.55,angle=-90]{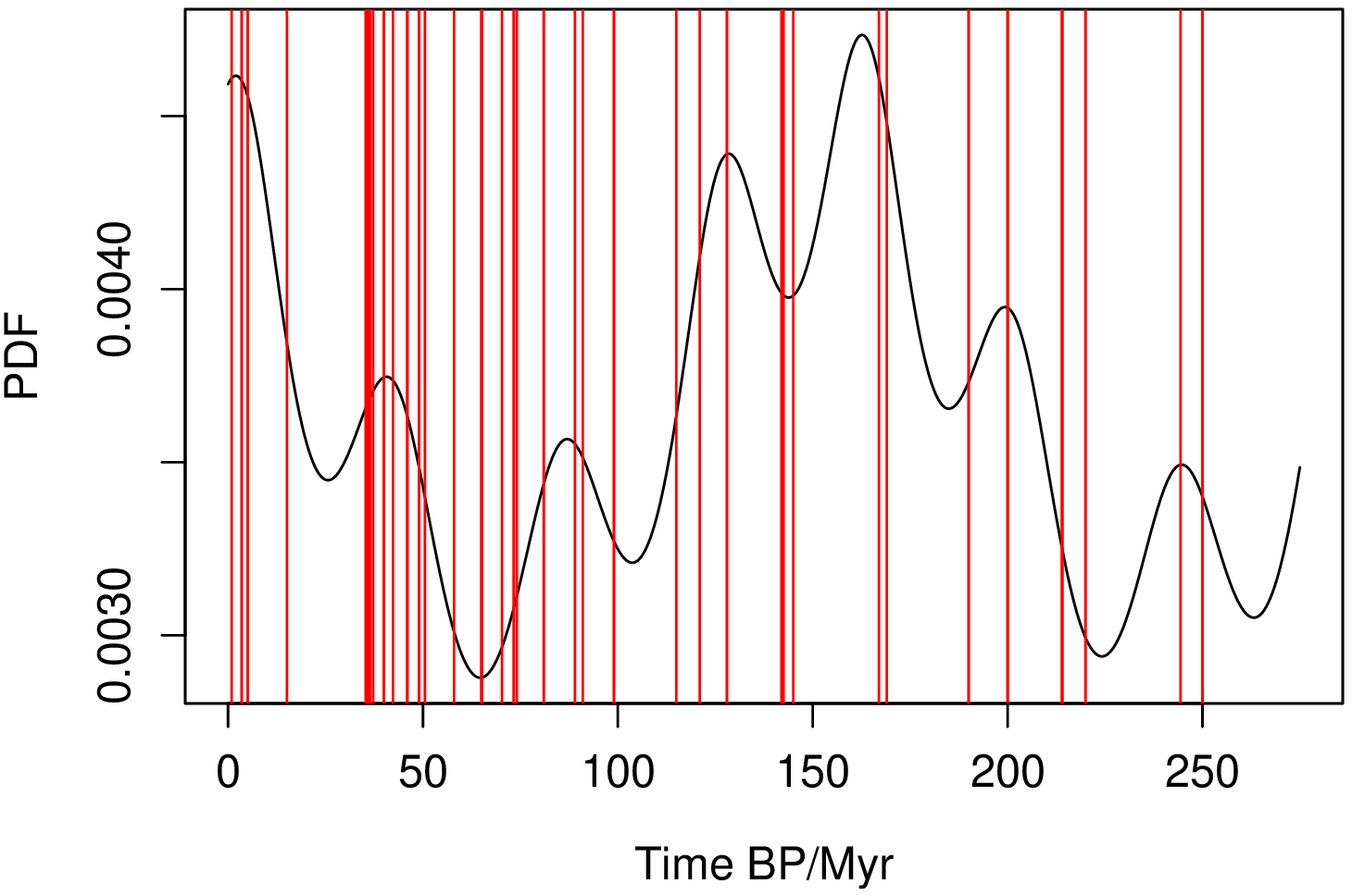}
  \includegraphics[scale=0.55,angle=-90]{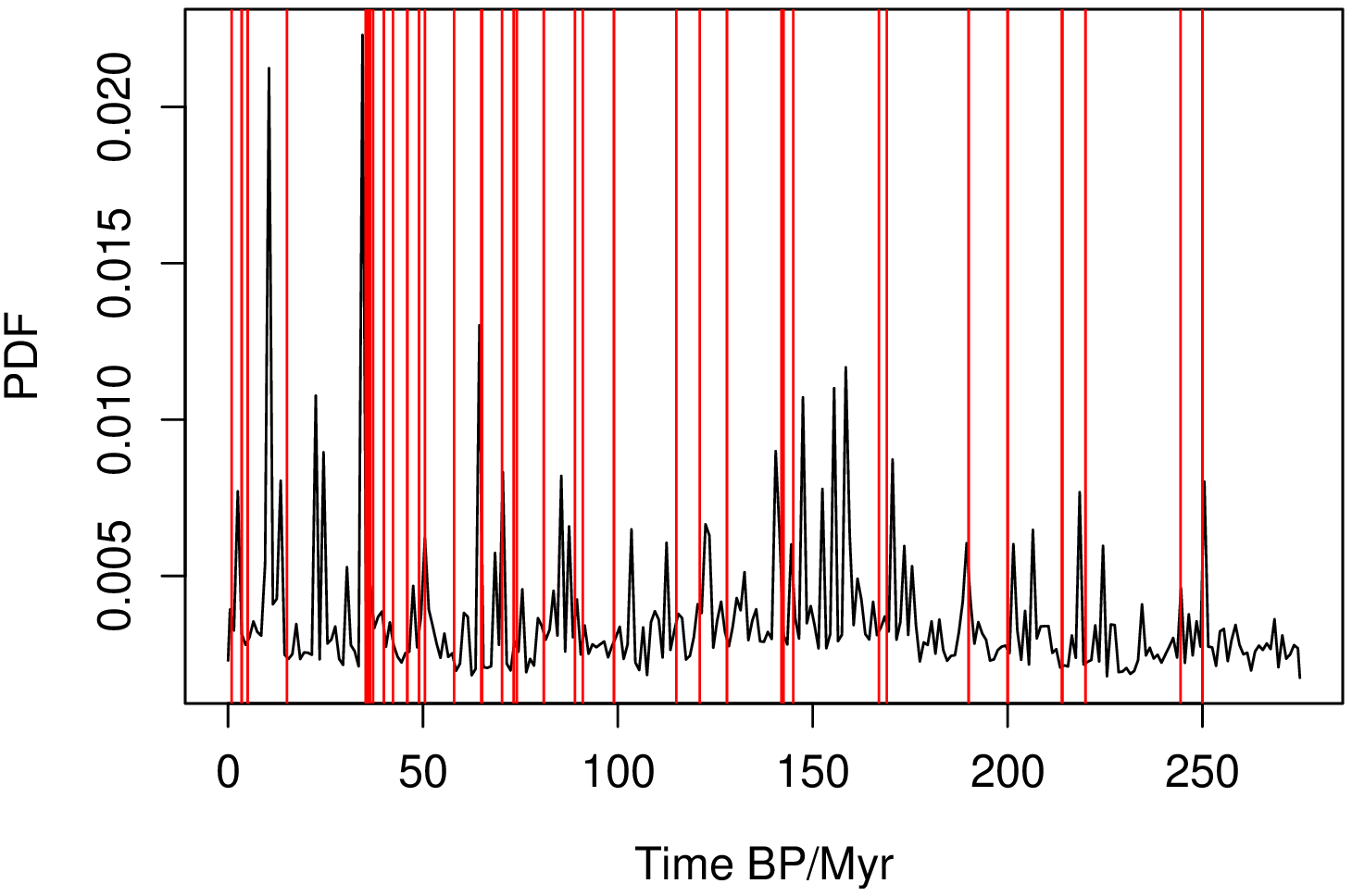}
  \includegraphics[scale=0.55,angle=-90]{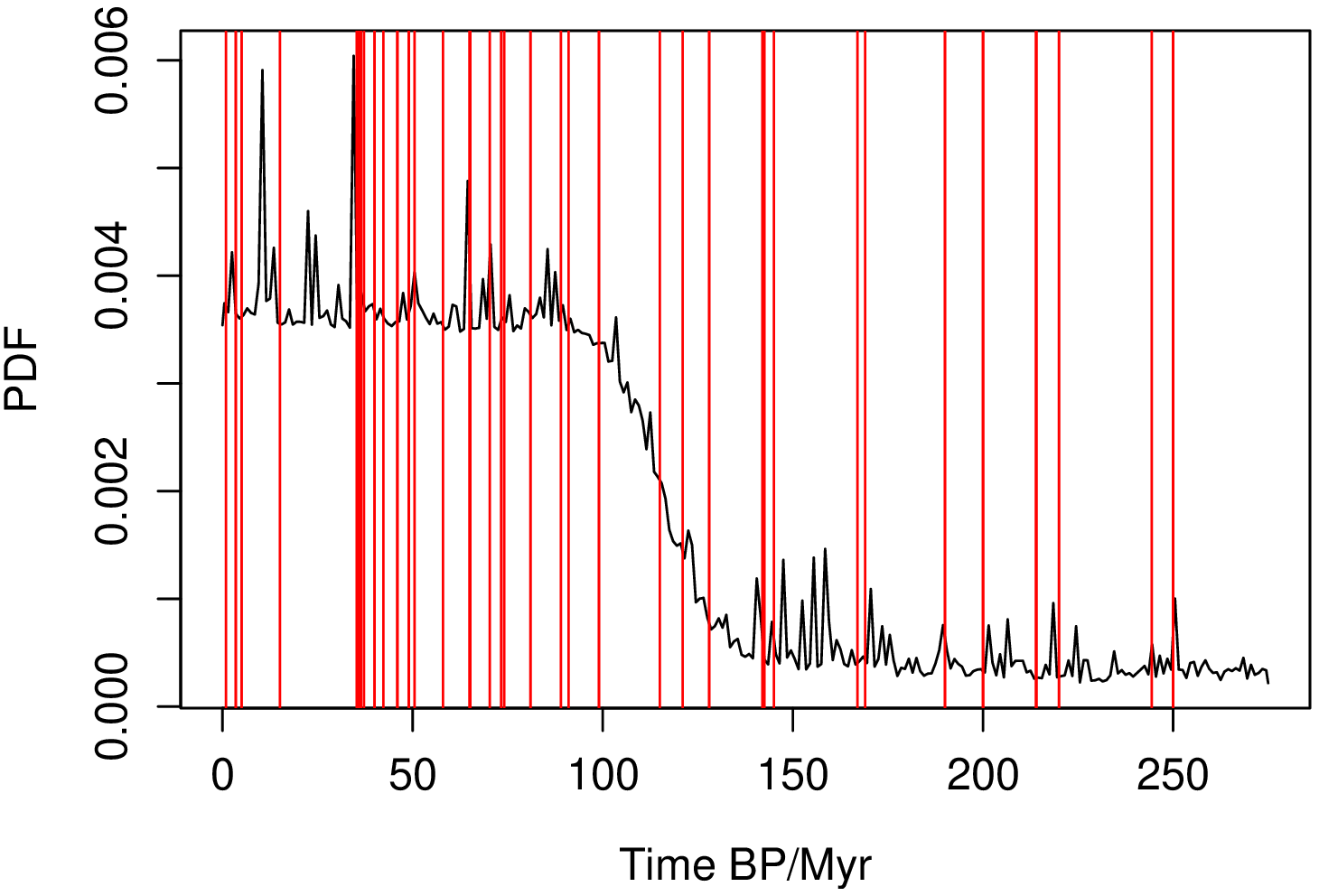}
  \caption{The prediction of the normalized cometary impact rate (i.e.\ probability density function; black line) compared to the actual impacts in the 
    basic250 time series (red lines). The models from top to bottom are TideProb, EncTideProb, and EncTideSigProb. A common solar orbit and encounter
    sample is used in all three cases.}
  \label{fig:dynmodels}
\end{figure}

Having defined TideProb and EncProb, we can combine them to make EncTideProb.
We can further combine this sum with SigProb (scaled by the parameter $\eta$) in order to include a smoothly varying component (see Table \ref{tab:tsmodels}).
Figure \ref{fig:dynmodels} shows examples of the TideProb, EncTideProb and
EncTideSigProb model predictions of the cometary flux for specific values of
their parameters. In the upper panel, we see the TideProb model predicts an
oscillating variation on at least two time scales. In the middle panel, we add
EncProb to TideProb. The amplitude of the background is reduced
due to the normalization effect -- the encounters dominate -- and the high
peaks characterize encounter-induced comet showers. In the bottom panel, the
SigProb model is added onto the EncTideProb model with $\eta=3$. A large value
of $\lambda$ has been used in SigProb here, such that the additional trend is
almost linear. Meanwhile, we also combine TideProb and SigProb to make TideSigProb. This of course does not show the randomly occurring peaks which are characteristic of the encounters model.

In Section \ref{sec:comparison}, we will compare these models with other time
series models defined in Section \ref{sec:tsmodel} using Bayesian method.

\section{Model comparison}\label{sec:comparison} 

Now that we have a way to generate predictions of the comet flux from our dynamical time series models, we use the Bayesian method described in section \ref{sec:bayesian} to calculate the evidences for the various time series models defined in section \ref{sec:tsmodel} for different cratering data sets. Because the solar orbit is more sensitive to the Sun's initial galactocentric distance ($R$) and angular velocity ($\dot{\phi}$) than to the other four initial conditions \citep{feng13}, we sample over only those two parameters when calculating the evidences and Bayes factors (ratio of two evidences) for the dynamical models. In order to make our model comparison complete, we will vary all initial conditions individually and simultaneously in section \ref{sec:sensitivity}.

To calculate the evidences we sample the parameter space of the dynamical
models and other time series models with $10^4$ and $10^5$ points
respectively. For the models of EncProb, EncTideProb, EncSigProb and
EncTideSigProb, each point represents an entire simulation of the orbit of the Sun about the Galaxy
and the corresponding simulation of the comet flux as a function of time. For
the latter we use the proxies of $G_3(t)$ and $\gamma(t)$ (i.e.\ the time-varying $\gamma_{\rm bin}$) described in section~\ref{sec:tideflux} and section~\ref{sec:encounterflux} respectively. For each orbit of the Sun we just generate a single sequence $\gamma(t)$ for the comet flux at random. (Because $\gamma(t)$ is modulated by the vertical tide coefficient $G_3(t)$, an average over many sequences of $\gamma(t)$ would be smooth and lack the spikes corresponding to comet showers which we see in the individual sequences.)

The Bayes factors of various models relative to the uniform model are listed
in Table \ref{tab:crater_BF}. We see that the SigProb, EncSigProb, TideSigProb and EncTideSigProb models are favoured by all the data sets, sometimes marginally, sometimes by a significant amount relative to certain models. In these favoured models, the negative trend (a decreasing cratering rate towards the past) is favoured much more than the positive trend. Such a negative trend can be picked out in Figure \ref{fig:set3lim}. As the positive values are so clearly ruled out, we only use negative values of $\lambda$ in all the trend models. This would be consistent with the crater preservation bias or the disruption of a large asteroid dominating over any recent increase in the asteroid impact rate (see section \ref{sec:impact}). 
\begin{table*}
  \centering
    \caption{Bayes factors of the various time series models (rows) relative to the uniform model for the various data sets (columns). The suffix numbers 1 and 2 in the model names, e.g.\ EncProb1 and EncProb2, refer to which different initial conditions are fixed. 1 means $R(t=0)$ and 2 means $\dot\phi(t=0)$.
}
    \begin{tabular}{@{} l|l l l l l l}
      \hline
      \hline
      Model      &basic150&ext150 &full150&basic250&ext250 &full250 \\
      \hline
      RandProb   &4.4     &9.3    &72    &3.0    &9.4   &4.7$\times 10^2$\\
      RandBkgProb&1.8     &3.8    &31    &2.2    &5.2&1.8$\times 10^2$\\ 
      SinProb    &0.34     &0.62    &1.2     &0.43    &0.76    &1.5     \\
      SinBkgProb &1.0     &1.2    &1.6    &1.0    &1.2    &1.5     \\
      SigProb    &15    &63   &$9.1\times 10^3$&$2.0\times 10^2$ &$1.8\times 10^3$&$5.8\times 10^6$\\
      SinSigProb &10     &36   &$1.6\times 10^2$&$1.0\times 10^2$&$6.0\times 10^2$&$2.6\times 10^5$\\
      EncProb1   &1.5     &3.9    &26     &1.7   &5.2&$1.1\times 10^2$\\
      EncProb2   &1.7     &3.3    &77    &1.6    &8.5&$2.7\times 10^2$\\
      TideProb1   &0.73   &0.87   &6.7   &0.81    &0.91&1.1\\
      TideProb2   &0.79   &0.86   &10    &0.69    &0.76&0.94\\
      EncTideProb1&1.0    &1.6   &18 &1.3 &2.1&10\\
      EncTideProb2&1.2    &1.8  &25   &1.2    &2.1&24\\
      EncSigProb1 &11     &41 &$4.6\times 10^3$&$1.5\times 10^2$&$1.5\times 10^3$&$5.9\times 10^6$\\
      EncSigProb2 &12     &52 &$8.7\times 10^3$&$1.7\times 10^2$&$1.5\times 10^3$&$6.6\times 10^6$\\
      TideSigProb1&11   & 38  &$4.6\times 10^3$&$1.6\times 10^2$&$1.4\times 10^3$&$6.2\times 10^6$\\
      TideSigProb2&10   & 37  &$4.5\times 10^3$&$1.6\times 10^2$&$1.4\times 10^3$&$6.1\times 10^6$\\
      EncTideSigProb1&11&40   &$5.0\times 10^3$&$1.6\times 10^2$&$1.4\times 10^3$&$6.0\times 10^6$\\
      EncTideSigProb2&11&40  &$4.7\times 10^3$&$1.6\times 10^2$&$1.5\times 10^3$&$6.1\times 10^6$\\
      \hline
    \end{tabular}
    \label{tab:crater_BF}
\end{table*}

The SinSigProb model is not favoured more than SigProb, which means the periodic component is not necessary in explaining cratering time series. This is consistent with the conclusion in \citet{bailer-jones11}. Moreover, the pure periodic model is actually slightly less favoured than the uniform model for the ``basic'' and ``ext'' data sets.  The pure random model (RandProb) is slightly more favoured than the random model with background (RandBkgProb). Both are more favoured than the uniform model, but with relatively low Bayes factors compared to the models with trend components.

EncProb is slightly more favoured than the TideProb model.
This suggests that the stochastic component of EncProb is slightly preferable to the smooth tidal component of TideProb in predicting the cratering data, although the difference is small.
Combining them to make the EncTideProb models does not increase the evidence.

The best overall model for explaining the data is SigProb, the pure trend model. Adding the tide or encounters or both does not increase the evidence by a significant amount for any of the data sets. This suggests that the solar motion has little influence on the total observed impact rate (i.e.\ comets plus asteroids and the preservation bias) either through the Galactic tide or through stellar encounters, at least not in the way in which we have modelled them here. This minor role of the solar motion in generating terrestrial craters weakens the hypothesis that the (semi-)periodic solar motion triggers mass extinctions on the Earth through modulating the impact rate, as some have suggested \citep{alvarez84,raup84}. We note that a low cometary impact rate relative to the asteroid impact rate has been found by other studies \citep{francis05,weissman07}.

The evidence is the prior-weighted average of the likelihood over the parameter space.  It is therefore possible that some parts of the parameter space are much more favoured than others (i.e.\ there is a large variation of the likelihood), and that this is not seen due to the averaging. In that case changing the prior, e.g.\ the range of the parameter space, could change the evidence. (We investigate this systematically in section~\ref{sec:sensitivity}).  In other words, the tide or encounter models may play a more (or less) significant role if we had good reason to narrow the parameter space. This would be appropriate if we had more accurate determinations of some of the model parameters, for example. We now investigate this by examining how the likelihood varies as a function of individual model parameters (but still be averaged over the other model parameters).

\begin{figure}
  \centering
  \hspace*{-8mm} \includegraphics[scale=0.48]{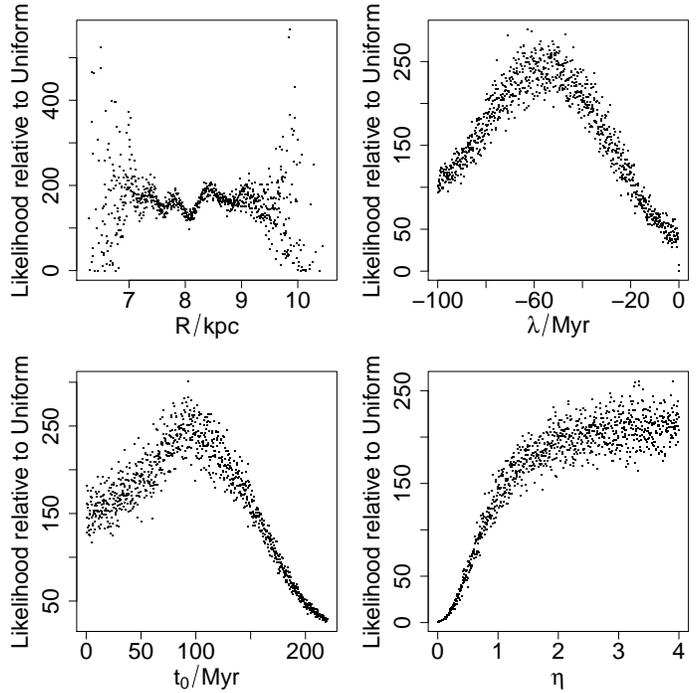}
  \caption{The distribution of the likelihood over each of the parameters in the TideSigProb1
    model for the basic250 data set, sampling over all other parameters in each case. 
    The parameters are divided into 1000 bins. For
    each bin, the likelihoods are averaged to reduce the noise generated by
    the randomly selected sequence of stellar encounters. There are 100\,000 samples in the parameter space. }
  \label{fig:like_1D}
\end{figure}

Figure \ref{fig:like_1D} shows how the resulting likelihood varies as a function of the
four parameters in the TideSigProb1 model.  The most favoured
parameters of the trend component are $\lambda \approx -60$\,Myr and $t_0
\approx 100$\,Myr. This trend component represents an increasing cratering
rate towards the present over the past 100 Myr
\citep{shoemaker98,gehrels94,mcewen97}, either real or a result of preservation bias. In the upper left graph, the
likelihood varies with $R$ slightly and varies a lot in the region where
$R<8$\,kpc and $R>9$\,kpc.  In the lower right panel, the likelihood increase with $\eta$,
which means that the trend component is important in increasing the likelihood
for the TideSigProb model. 

To find the relationship between the likelihood for TideSigProb and the
Sun's initial galactocentric distance $R$ and the scale parameter $\eta$, we
fix the parameters of the trend component to $\lambda=-60$\,Myr and
$t_0=100$\,Myr.
\begin{figure}
  \centering
  \hspace*{-4mm}\includegraphics[scale=0.75]{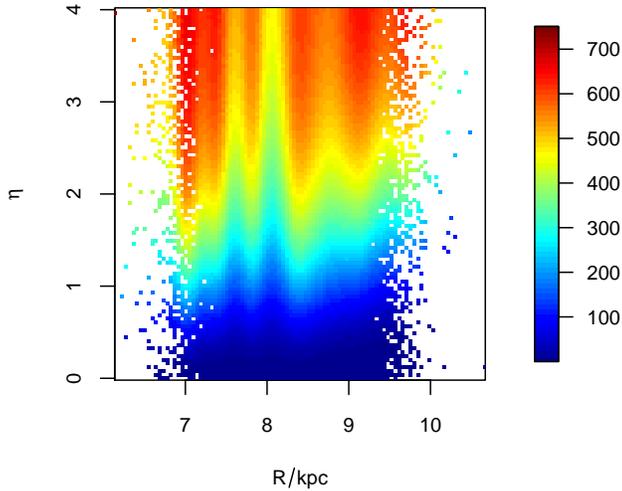}
  \caption{The distribution of the likelihood over the parameters $R$ and $\eta$ in the TideSigProb1 model relative to the Uniform model for the  basic250 data set. The relative likelihood is shown as the colour scale indicated in the legend. There are 100\,000 samples in the parameter space. }
  \label{fig:like_2D}
\end{figure}
In Figure \ref{fig:like_2D} we see that the likelihood for TideSigProb
increases monotonically with $\eta$ over this range, but has a more complex dependence on $R$. 
The likelihood is highest at around $R=7.0$ and $R=9.5$\,kpc.
In Figure \ref{fig:tideprob_R70} we compare the dates of the craters in the basic250 data set
with the prediction of the cratering rate from TideProb with $R=7.0$\,kpc.
There are 7 craters within the first 30\,Myr compared to 16 and 13 craters in
the intervals [30,60]\,Myr and [60,90]\,Myr respectively. This lack of craters
in the first 30\,Myr can be better predicted by TideSigProb than by
the SigProb model with a negative $\lambda$. 
While this is small number statistics, it may suggest that even though we have little evidence for
the effect of the tide on cometary impacts in the overall cratering data, it may have had more of an effect in selected time periods. Other explanations are also possible, of course: we cannot say anything about models we have not actually tested, such as a more complex model for the asteroid impact rate variation.
\begin{figure}
  \centering
  \includegraphics[scale=0.5,angle=-90]{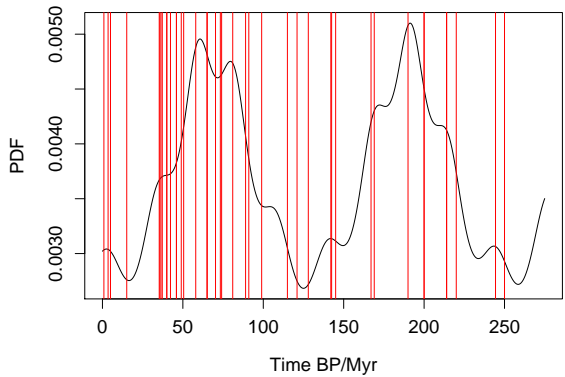}
  \caption{Comparison between the prediction of TideProb with $R= 7.0$\,kpc (shown as a probability distribution function in black) and the times of the impact craters in the basic250 data set (shows as vertical red lines). } 
  \label{fig:tideprob_R70}
\end{figure}

\section{Modelling the angular distribution of cometary perihelia}\label{sec:ADP}

In this section we predict the 2D angular distribution (latitude, longitude) of the perihelia of LPCs, the observed data for which are shown in Figure \ref{fig:LPC1A_bl}.  To do this we need to identify from the simulations comets injected over an appropriate time scale.  Figure \ref{fig:gamma_fc_pdf} shows that a comet shower usually has a duration of less than 10\,Myr, something which was also demonstrated by \cite{dybczynski02} in detailed simulations of individual encounters.  The Galactic tide varies little over such a time scale, because the vertical component of the tide, which dominates the total Galactic tide, varies over the period of the orbit of the Sun about the Galaxy, which is of order 200\,Myr.  We may therefore assume that the solar apex is also more or less fixed during the past 10\,Myr, which is then an appropriate time scale for constructing our sample. 

We simulate cometary orbits over the past 10\,Myr as follows: (1) generate one million comets from the Oort cloud model (DLDW or DQT), as well as a set of stellar encounters (about 400 over 10\,Myr); (2) integrate the cometary orbits under the perturbations of only the Galactic tide (tide-only simulations with a time step of 0.1\,Myr), only stellar encounters (encounter-only simulations with a time step of 0.01\,Myr), and both of them (combined simulations with a time step of 0.01\,Myr) back to 10\,Myr ago; (3) identify the injected comets and their longitudes and latitudes. We then repeat steps (1)--(3) ten times (i.e.\ resample the Oort cloud and the set of stellar encounters) and combine the results in order to increase the number statistics.

\subsection{Latitude distribution}\label{sec:latitude}

\begin{figure}
  \centering
  \vspace*{2in}
\hspace*{-1em}\includegraphics[width=0.5\textwidth]{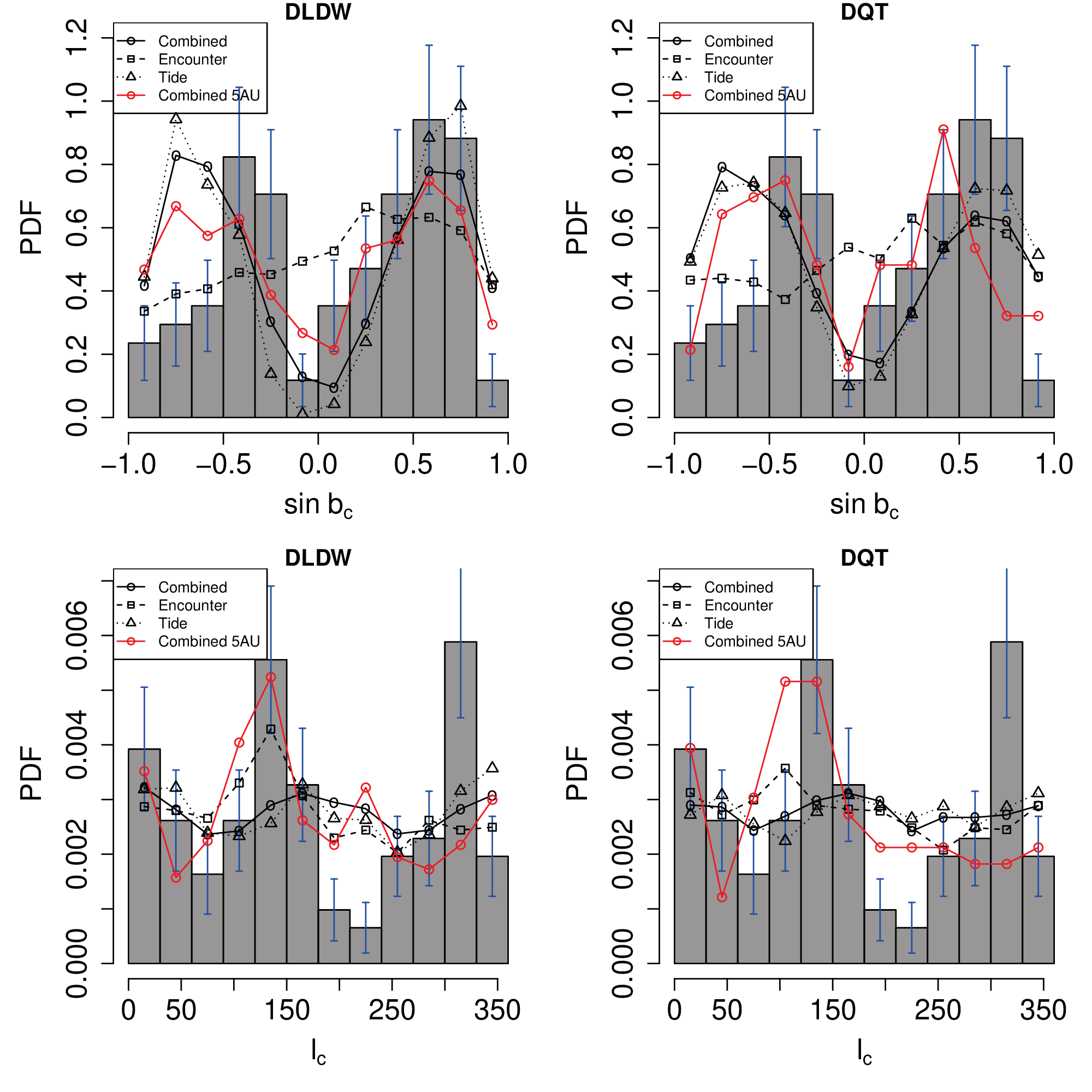}
  \vspace*{-1.60in}
\caption{Comparison between the observed distribution (histogram blocks) and model-predicted distributions (points/lines) of the perihelia of long-period comets (LPCs) with Galactic latitude (upper panels) and longitude (lower panels) for the DLDW (left panels) and the DQT (right panels) Oort cloud initial conditions.  All distributions are normalized. The error bars on the data have been calculated using a Poisson noise model (arising from the binning) with a total of 102 class 1A LPCs. The model-predicted distributions show the comets injected into the loss cone for three modes of simulations, namely including only the Galactic tide (triangles), only stellar encounters (squares), and both (circles). The number of injected comets in these simulations for the DLDW (DQT) models are 1858 (981), 1133 (1976), and 12\,751 (2796), respectively.  The red circles connected by red lines show the number of comets injected into the observable zone ($q<5$\,AU), and comprise 449 comets for the DLDW model, and 112 for the DQT model. }
  \label{fig:dyn_bl}
\end{figure}

The upper panels of Figure \ref{fig:dyn_bl} compare the Galactic latitudes of the LPC perihelia with our model predictions. In addition to showing the model predictions for the comets injected into the loss cone, we also show the predicted distributions for comets injected into the observable zone ($q<5$\,AU). The former contains more comets, but the latter is of course closer to the observed sample.
The small sample of comets within the observable zone have significant sample noise in their angular distributions, so we will only compare model predictions of the angular distribution of comets in the (larger) loss cone.

The upper panels show that the injected LPCs in the pole and equatorial regions are depleted for both DLDW and DQT models, as also found by \cite{delsemme87}.  According to theoretical prediction, the tide-induced flux should be proportional to $|\sin b \cos b|$ \citep{matese99},
in very good agreement with our tide-only simulations.  The observed data broadly agree with this, the main difference being that for negative latitudes the peak is at around -0.4 rather than the model-predicted value of -0.7. This discrepancy was also noticed by \citet{matese11}, for example, and could be a consequence of the small size of the data set (note the errors bars in the figure).

We see in the figure that the PDF of the latitude distribution predicted by the combined simulation always lies between those predicted by the single perturbation simulations.
Although the combined simulation of comets injected into the loss cone predicts a flatter distribution than the tide-only simulation does, the stellar encounters cannot entirely smooth out the peaks in the latitude distribution. This is consistent with the results in \cite{rickman08}. Thus the observed non-uniform latitude distribution does not indicate that the Galactic tide dominates at the present epoch, as was claimed by \cite{matese11}.

We can attempt to make a more quantitative assessment of how well our models predict the observed distribution. Using model comparison techniques we can ask whether our dynamical models (the combined tide plus encounters model) explain the data better than a uniform distribution. 
We can do this crudely on the binned data/simulations shown in the figure via a likelihood test. The act of binning means that the model-predicted number of events per bin is determined by the Poisson distribution, thus defining our likelihood. However, such a test is dependent on the choice of binning, and we have tried out a range of bin widths and centres. While we find that the combined model for the DQT Oort cloud model is always more favoured than a uniform distribution, the significance is marginal.

An alternative approach is to use the unbinned data and unbinned model predictions, and to apply a kernel density estimate (KDE) to each. This produces a non-parametric density function for the data and for the model, the difference between which we quantify using the (symmetrized) Kullback-Leibler divergence (KLD). A value of zero divergence means that the two distributions are identical; larger (positive/negative) values indicate larger differences.  We find that our dynamical models give smaller KLD values than do the uniform model (i.e.\ the former predict the data better), for both the DLDW and DQT. Although the distributions formed by the KDE are sensitive to size of the kernel adopted,\footnote{This is analogous to the size of the histogram bins. A histogram is just a particular type of kernel.}  we find that the KLD values are quite insensitive to this, and consistently favour the dynamical models.  This suggests that the dynamical models explain the data better than a flat distribution in latitude (although because calibrating KLD ratios into formal significances is not easy, we leave this as a qualitative statement).

\subsection{Longitude distribution}\label{sec:longitude}

The perihelia of LPCs are not distributed uniformly on the celestial sphere. It has been suggested \citep{matese99, matese11} that they lie preferentially on a great circle, as evidenced by two peaks at $l_c\simeq135$\deg and $l_c\simeq315$\deg seen in Figure \ref{fig:LPC1A_bl}.  The comets on this great circle could be induced by stellar encounters with preferred directions, thereby producing the apparent anisotropy.  In the lower two panels in Figure \ref{fig:dyn_bl}, we see that the model predictions do not produce any very large peaks, although one around $l_c\simeq135$\deg\ is discernable.  We also observe a peak around $l_c =$\,0--60\deg\ which is proposed as a signal of the ``Biermann comet shower'' \citep{biermann83,matese99}. In our model, this peak is probably the result of accumulated perturbations from several stellar encounters with preferred directions.
  
The peak around $l_c=135$\deg\ is more prominent in the model prediction for the comets injected into the observable zone (red points/line in the figure). This peak is generated primarily by one or more massive stellar encounters. Hence, stellar encounters play a more significant role in injecting comets into the observable zone than just into the loss cone. This is consistent with the ``synergy effect'' investigated by \cite{rickman08}. 

As with the latitude distribution, we also measured the KLD for the model predictions (for the loss cone) and for a uniform distribution. The dynamical models predict the data little better than a uniform distribution. (The likelihood test gives a similar result.) One reason for this lack of support for our dynamical (combined) model could be the fact that we are averaging the predicted distribution from the encounters over ten different realizations of the stellar encounters. This will tend to smooth out individual peaks, which are probably produced by just a few encounters with massive stars.\footnote{Such massive stars (or stars with relatively high $\gamma$) move slowly relative to the Sun, and so would generate a relatively narrow peak in comet flux with $l_c$.}
If we instead only used a single random realization of encounters, we are unlikely to reproduce exactly the showers which occurred. This is an inherent problem of modelling stellar encounters in a stochastic way.  This does not affect our model prediction of the latitude distribution nearly as much, however, because its shape is dominated by the non-stochastic tide.

In order to investigate this we again use our encounter model via the proxy $\gamma$ (a proxy of comet flux) defined in equation \ref{eqn:gamma}, but now as a function of $b_{\rm p}$ and $l_{\rm p}$, the direction toward the perihelion of the stellar encounter. Moreover, we now impose a minimum threshold, $\gamma_{\rm lim}$, on the proxy: The larger the value of $\gamma_{\rm lim}$, the larger the encounter perturbation must be for it to be included in the model.

\begin{figure}
  \centering
  \hspace*{-4mm}\includegraphics[scale=0.65,angle=-90]{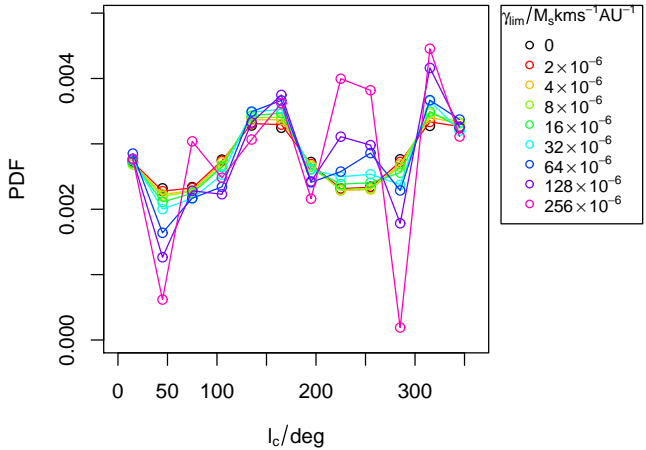}
  \caption{Predictions of the enounter-induced cometary flux when adopting different lower limits, $\gamma_{\rm lim}$, on the value of $\gamma$ required for an event to have an influence on the Oort cloud. There are $10^7$ and $10^8$ encounters generated for the model predictions with $\gamma_{\rm lim}=0$ and $\gamma_{\rm lim}\neq0$ respectively. 
}
  \label{fig:l_gamma_lim}
\end{figure}

Using the encounter model described in section \ref{sec:encmod}, we simulate 10 million encounters and calculate $\gamma$, $b_{\rm p}$, and $l_{\rm p}$ for each. The predicted direction of an LPC's perihelion is opposite on the sky to the direction of the encounter perihelion. Thus we can calculate $b_c$ and $l_c$ accordingly and use $\gamma(b_c,~l_c)$ to predict the PDF of $b_c$ and $l_c$.  Then we divide the range of the Galactic longitude into 12 bins and sum $\gamma$ in each bin including only those encounters with $\gamma>\gamma_{\rm lim}$. Normalizing this gives the angular PDF of the encounter-induced flux, as shown in Figure \ref{fig:l_gamma_lim}. For larger values of $\gamma_{\rm lim}$ we observe a larger variation in the flux with longitude, as expected, because then fewer encounters contribute to the distribution. As we can see from equation \ref{eqn:gamma}, these are the more massive and/or slower stars. These encounters may induce a series of weak comet showers rather than a single strong comet shower. Because strong encounters are rare and extremely weak encounters cannot induce enough anisotropic LPCs, the spikes in the longitude distribution can be caused by at least two weak encounters rather than one strong or many extremely weak encounters. From Figure \ref{fig:dyn_bl}, we see that the tide cannot completely wash out the anisotropy in the longitude distribution induced by these encounters. 

Consistent with our results, \cite{matese11} found that the two spikes in the longitude distribution result from weak impulsive perturbations by analyzing the energy and angular momentum of dynamically new LPCs. Similar to the definition of weak comet showers in \cite{matese02} and \cite{dybczynski02}, we define encounters with $\gamma$ in the interval $\lbrack 1\times 10^{-7}, 5\times 10^{-6}\rbrack M_{\sun} ~km~s^{-1}~AU^{-1}$ as weak encounters. We do not find strong peaks in the longitude distribution of $\gamma$ for these encounters in Figure \ref{fig:l_gamma_lim},  because we know that $\gamma$ can underestimate the intensity of the shower (see Figure \ref{fig:gamma_fc_pdf}). Thus a small enhancement of the two peaks in Figure \ref{fig:l_gamma_lim} may correspond to a large enhancement of the peaks in the longitude distribution as predicted by our dynamical model in Figure \ref{fig:dyn_bl}.

Inspecting the catalogue of the frequencies of different types of stellar encounters in table 8 of \cite{sanchez01}, we see that there were at least eight encounters with masses equal to or larger than one solar mass encountering the solar system in the past 10\,Myr with perihelia less than 1\,pc. These encounters can move to a heliocentric distance much larger than 50\,pc over that time, which is the upper limit for their unbiased sample of stellar encounters with $M_V < 5$ -- see Figure 13 of \cite{sanchez01}. 

We also point out that GL 710 will have a close approach with the solar system in about 1.4\,Myr at a perihelion longitude of around 135$^\circ$. According to studies, it will induce a weak comet shower which is expected to increase the cometary flux by 40\%-50\% \citep{sanchez99,matese02}. This supports the suggestion that the solar apex motion induces the non-uniform longitude distribution of the LPCs' perihelia (see Figure \ref{fig:venc_denc_bl} and \ref{fig:dyn_bl}). In addition, Algol, a triple-star system with a total mass of 5.8\,$M_{\sun}$, encountered the solar system with a closest distance of 2.5\,pc 6.9\,Myr ago \citep{sanchez01}. The Galactic longitude of Algol was also close to $135^\circ$.

Based on the above plausible scenario, we conclude that the peaks in the longitude distribution of LPC perihelia could arise from the perturbations of a few strong stellar encounters, the encounter directions of which depend on the solar apex motion. Considering the important role of the Galactic tide in generating a non-uniform latitude distribution, and the role of stellar encounters in generating a non-uniform longitude distribution, the synergy effect plays a role in maintaining -- rather than smoothing out -- the anisotropy in the observed LPCs. In other words, we can explain the anisotropy of the LPC perihelia based only on the solar apex motion and the Galactic tide, without needing to invoke the Jupiter-mass solar companion as proposed by \cite{matese11}. To date there is no observational evidence for such a companion. We note that a recent analysis of data from the WISE satellite has excluded the existence of a Jupiter-mass solar companion with a heliocentric distance less than 1\,pc \citep{luhman14}.

\section{Sensitivity test}\label{sec:sensitivity}

\subsection{Spiral arms and Galactic bar }\label{sec:armbar}

The spiral arms and Galactic bar are non-axisymmetric, time-varying components of the Galactic potential. These make only a small contribution to the tidal force acting on the Sun and Oort cloud (\citet{binney08_book,cox02}). However, if their contribution is always in the same direction, the effect of their perturbation could accumulate.  This can occur when the Sun is near to the co-rotation resonance, when the rotation velocities of the disk and of the spiral pattern coincide.  To test this hypothesis, we simulate the solar and cometary motion adopting various constant pattern speeds of the spiral arms and the bar with fixed Galactic density distributions (specified in Section \ref{sec:potential}).

\begin{figure}
  \centering
  \hspace{-2mm}\includegraphics[scale=0.43]{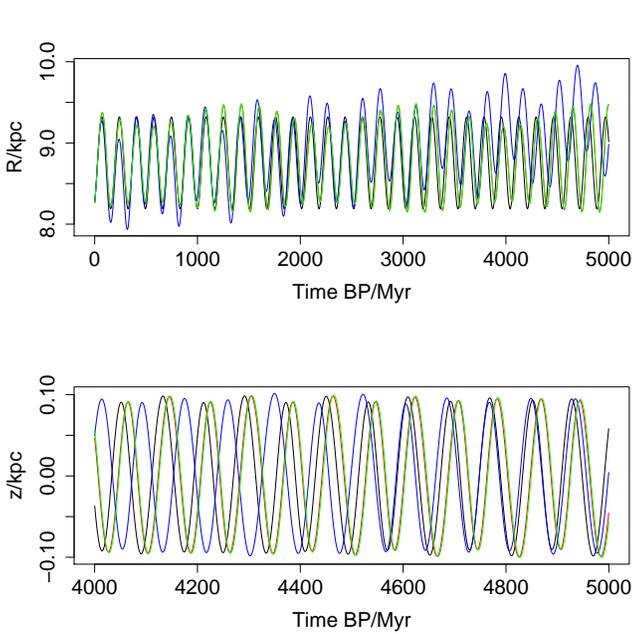}
  \caption{The variation of Sun's galactocentric radius (upper panel) and
    vertical displacement from the disk (lower panel) as calculate for
    different potentials: axisymmetric potential (black); potential including
    Galactic bar (red); potential including spiral arm (blue); potential
    including both bar and arm (green). To show different lines in the lower
    panel better, we plot the variation of the Sun's vertical displacement
    over a shorter time scale.} 
  \label{fig:solar_orbit}
\end{figure}

We integrate the solar orbit in the Galactic potential both including and excluding the non-axisymmetric components. The initial conditions of the Sun and potential parameters are given in Table \ref{tab:model_par}. We find that the gravitational force from the bar is always much larger than that from the spiral arms. However, the difference between the pattern speed of the Galactic bar $\Omega_b$ and solar angular velocity is much larger than the difference between the pattern speed of the spiral arms $\Omega_s$ and solar angular velocity, which results in a much lower accumulated perturbation due to the bar. To see this effect, we integrate the solar orbit back to 5 Gyr in the past.  The variations of galactocentric radius and vertical displacement of the Sun are shown in Figure \ref{fig:solar_orbit}.  The arms have a stronger effect on the solar orbit than does the bar.  The spiral arms tend to increase the galactocentric radius of the Sun as the integration proceeds (back in time), while the bar modulates the galactocentric radius by a comparatively small amount. Neither the bar nor the arms significantly affect the vertical displacement amplitude of the Sun. Here the combined perturbation from the potential including both the Galactic bar and spiral arms changes the solar motion the same way as the perturbation from the bar alone.  

\begin{figure}
  \centering
  \includegraphics[scale=0.8,angle=-90]{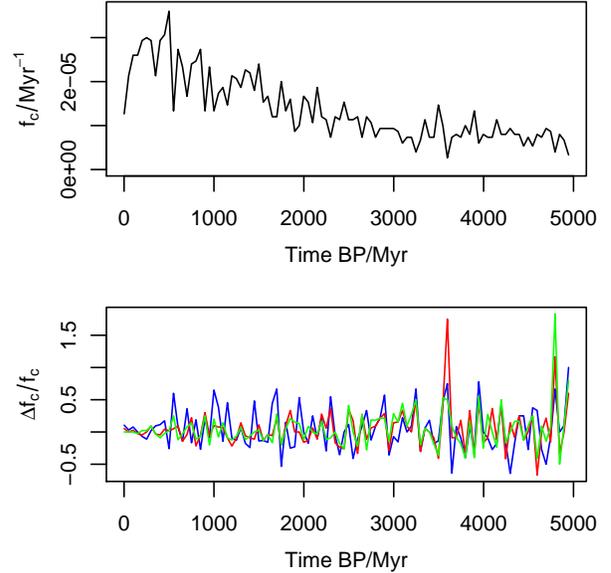}
  \caption{The magnitude of the tide-induced flux, $f_c$, generated by the
    axisymmetric potential model (upper panel) and the relative flux difference,
    $\Delta f_c/f_c$, generated by asymmetric Galactic potential models (lower panel) over the past 5 Gyr with a sample of 3$\times 10^4$ comets. The potentials are: axisymmetric potential only (black); including the arms (blue); including the Galactic bar (red); including both the arms and the Galactic bar (green).
} 
  \label{fig:flux_nonsym}
\end{figure}

We now simulate the tide-induced flux corresponding to these different
potential models. The lower panel in Figure \ref{fig:flux_nonsym} shows that the non-axisymmetric components do not alter the flux very much. Although the
perturbation from the arms can change the solar orbit slightly, the resulting change in the perturbation of the Oort cloud is minimal. The changed tidal force may change some individual cometary orbits, but has little effect on the overall injected comet flux, because the effect of the tide depends also on the distribution of the comets, which is nearly isotropic. 
We also see that the arms modify the cometary flux more than the bar,
consistent with its larger impact on the stellar density. (The limited number
of injected comets contributes to the sharp peaks in the relative flux difference, $\Delta f_c/f_c$, after 3\,Gyr.)

We also investigated the sensitivity of the solar motion and comet flux to the pattern speed of the asymmetric components.  We find that the closer the pattern speed of the arms is to the angular velocity of the Sun, the larger the perturbation from the arms is. (We can understand this in terms of a resonance.) Meanwhile, the perturbation from the bar is not sensitive to the bar's pattern speed.

Finally, we also find that the distribution of $b_c$ and $l_c$ of the comet flux does not change very much for different non-axisymmetric components of the Galactic potential.

In summary, we find that the model predictions of the tide-induced cometary flux are generally insensitive to changes in the non-axisymmetric components of the Galactic potential, except when a resonance between the arms and the solar orbit occurs, which increases the variation in the cometary flux.

\subsection{Variations of the prior}\label{sec:components} 

As discussed earlier, the evidence depends on the prior distribution adopted
for the model parameters. As this prior frequently cannot be determined with
any certainty, it is important to investigate the sensitivity of the evidence
to changes in the prior.\footnote{A more robust -- but also more
  time-consuming -- way of calculating the evidence is presented in
  \cite{bailer-jones12}.}  To complete the calculation of evidences for
dynamical models, we also vary the other three initial conditions,
$V_R(t=0~{\rm Myr})$, $z(t=0~{\rm Myr})$, and $V_z(t=0~{\rm Myr})$, in the
EncTideSigProb models, which we previously kept constant.  Together with
SigProb, EncSigProb and TideSigProb, this was previously the best favoured model (Table \ref{tab:crater_BF}).  We made numerous changes in the priors by altering their parameter ranges, and re-did all necessary Monte Carlo samplings, numerical simulations, and likelihood calculations and recomputed the Bayes factors. Some of our results are shown in Table \ref{tab:prior_change}.

\begin{table*}
\centering
\caption{The Bayes factors for various time series models (rows) relative to the uniform model for two different data sets (cf.\ Table \ref{tab:crater_BF}). The second column describes what change has been made to the range of which parameter in the prior. The other priors are kept fixed.
TideSigProb3--6 refer to the TideSigProb model in which different initial conditions are varied:
  $V_R(t=0~{\rm Myr})$; $z(t=0~{\rm Myr})$; $V_z(t=0~{\rm Myr})$; all three (respectively)}
\begin{tabular}{l|l|c|c}
\hline
\hline
models&varied prior&Bayes factor for basic150&Bayes factor for basic250\\
\hline
\multirow{5}{*}{RandProb}
&none&4.4&3.0\\
&$\sigma=2\bar{\sigma_i}$&2.0&4.8\\
&$\sigma=1/2\bar{\sigma_i}$&2.2&4.7\\
&$N=2N_{\rm ts}$&1.9&1.8\\
&$N=1/2N_{\rm ts}$&2.4&7.6\\
\hline
\multirow{5}{*}{RandBkgProb}
&none&1.8&2.2\\
&$\sigma=2\bar{\sigma_i}$&1.6&3.7\\
&$\sigma=1/2\bar{\sigma_i}$&1.8&2.6\\
&$N=2N_{\rm ts}$&1.5&1.5\\
&$N=1/2N_{\rm ts}$&2.4&2.9\\
\hline
\multirow{4}{*}{SinProb}
&none&0.34&0.43\\
&$10<T<100$&0.12&0.14\\
&$2\pi/300<\omega<2\pi/10$&0.34&0.39\\
&$10<T<300$&0.88&$5.4\times 10^{-2}$\\
\hline
\multirow{4}{*}{SinBkgProb}
&none&1.0&1.0\\
&$10<T<100$&0.90&0.88\\
&$2\pi/300<\omega<2\pi/10$&1.0&1.0\\
&$10<T<300$&1.8&1.4\\
\hline
\multirow{4}{*}{SigProb}
&none&15&$2.0\times 10^2$\\
&$0<t_0<1.2\tau_{\rm max}$&13&$1.4\times 10^2$\\
&$-100<\lambda<100$&7.7&$1.0\times 10^2$\\
&$0<\lambda<100$&$1.3\times 10^{-2}$&$1.8\times 10^{-3}$\\
\hline
\multirow{3}{*}{SinSigProb}
&none&6.4&80\\
&$0<t_0<1.2\tau_{\rm max}$&8.3&71\\
&$2\pi/300<\omega<2\pi/10$&9.9&97\\
\hline
TideSigProb3&none&9.0&$1.7\times 10^2$\\
\hline
TideSigProb4&none&9.1&$1.7\times 10^2$\\
\hline
TideSigProb5&none&9.0&$1.7\times 10^2$\\
\hline
TideSigProb6&none&11&$1.6\times 10^2$\\
\hline
\end{tabular}
\label{tab:prior_change}
\end{table*}

The difference in Bayes factors for random models (RandProb, RandBkgProb) and periodic models (SinProb, SinBkgProb) with different prior distributions is less than five. The Bayes factors also remain less than ten so they remain no better explanations of the cratering data than the Uniform model. Thus our former conclusions about these models are not very sensitive to plausible changes in the priors. 

The TideSigProb models in which other parameters are varied have nearly the same evidences as the TideSigProb models listed in Table \ref{tab:crater_BF}, so these too are insensitive to these changes in the priors.
We also see that the SigProb model with positive $\lambda$ has Bayes factors much lower than
SigProb with negative $\lambda$ for both the basic150 and basic250 data sets. 

The dynamical models have parameters of the Galaxy potential,
Sun's initial conditions and combination ratio parameters ($\eta$ and $\xi$) which are listed in Table \ref{tab:prior}). To keep things simple, we change the fixed parameters and the ranges of the
varying parameters individually, and then calculate the evidence by sampling
the prior defined by the changed parameter and other parameters shown in Table \ref{tab:prior}. We calculate evidences for dynamical models with double or half the disk mass ($M_d$), halo mass ($M_h$), standard deviation of the initial value $R$ ($\sigma_R$), and the range of the varying ratio between the EncTideProb (or TideProb) and SigProb models ($\eta$). In addition, previous studies suggest that the number of tide-induced LPCs is not identical to the encounter-induced LPCs, i.e.\ $\xi\neq1$ \citep{heisler87,rickman08}. Thus we multiply the ratio between the tide-induced flux and the encounter-induced flux ($\xi$) by a factor of 4 or 1/4 for the sensitivity test. 

The resulting Bayes factors calculated for the basic150 data set are
shown in Table \ref{tab:dyn_prior_change}. In each row we see little variation: the Bayes factors
are relatively insensitive to these parameters. This means that either the parameter space of the EncTideSigProb1 model is evenly favoured by the
basic150 data set, or the data are unable to discriminate between the compound dynamical models.
\begin{table*}
\centering
\caption{The Bayes factors for EncProb1, EncTideProb1 and EncTideProb1 for
  basic150 with different Galaxy parameters.
  }
\begin{tabular}{l|*{11}{c}}
\hline
\hline
models &none&$2M_d$&$1/2M_d$&$2M_h$&$1/2M_h$&$2\sigma_{R}$&$1/2\sigma_{R}$&$\xi=4$&$\xi=1/4$&$0<\eta<8$&$0<\eta<2$\\
\hline
\multirow{1}{*}{EncProb1}
&1.5&2.5&3.4&2.5&4.1&2.3&2.6&---&---&---&---\\
\hline
\multirow{1}{*}{EncTideProb1}
&1.0&2.1&2.3&2.6&3.5&1.8&1.0&1.5&0.73&---&---\\
\hline
\multirow{1}{*}{EncTideSigProb1}
&11&15&11&13&12&12&11&12&10&13&8.8\\
\hline
\end{tabular}
\label{tab:dyn_prior_change}
\end{table*}

The model prediction of the anisotropic LPCs (see Figure \ref{fig:dyn_bl}) depends to a greater or lesser extent on the Galactic potential, the Sun's initial condition, the Oort Cloud model, and the model of encounters. We vary the model parameters in the same way as we did in Table \ref{tab:dyn_prior_change} and simulate ten million orbits of DLDW comets perturbed by the tide and ten samples of stellar encounters backwards to 10\,Myr ago. We find that the latitude distribution of the LPC perihelia is not sensitive to the change of the Galactic halo mass, the initial conditions of the Sun, or the direction of the solar apex. The amplitudes of the peaks in the latitude distribution are reduced if we decrease the mass of the Galactic disk or increase the stellar masses, which make the stellar encounters play a more important role in injecting comets into the loss cone.
However, the overall profile of the peaks is not changed in the latitude distribution.

The peaks in the longitude distribution shift slightly if we change the solar apex direction, the masses of the encounters, or the mass of the Galactic disk. The longitude distribution is not sensitive to changes in the other model parameters.

Finally, we also tested the effect of changing the time step in the (combined) simulations. We simulated four million comets generated from the DLDW model perturbed by the tide and ten samples of stellar encounters backwards to 10\,Myr ago using a time step of 0.001\,Myr (as opposed to 0.01\,Myr). We find little change in either the latitude or longitude distributions. In addition, we see only 4\% more comets injected when using this smaller time step.
%We generate stellar encounters with half and double the solar apex velocity. We find that the higher the apex velocity, the less the variation of LPC flux with longitude. The longitude at which the flux is maximum of course varies with the direction of the solar apex. However, given the robust measurements of the solar apex velocity and direction \citep{sanchez01}, the overall distribution of the longitude is unlikely to vary significantly.

In summary, we find that the overall shape of the angular distribution of LPC perihelia in both longitude and latitude is not very sensitive to changes in the model parameters, in particular not to the initial distribution of Oort Cloud comets, not to the masses of Galactic halo and disk, and not to the initial conditions of the Sun.

\section{Discussion and Conclusion}\label{sec:conclusion}

We have built dynamical models for the impact rate and angular distribution of comets induced by the Galactic tide and stellar encounters, as modulated by the solar motion around the Galaxy.  Without using the approximate methods (the averaged Hamiltonian or impulse approximation), we numerically simulate the tide-induced flux and encounter-induced flux separately. We use these to validate the use of proxies for tide-induced flux, $G_3$, and for the encounter-induced flux, $\gamma_{\rm bin}$, in our models.  

Using the Bayesian evidence framework, we find that the pure trend model
(SigProb) together with the dynamical models including a trend component
(EncSigProb, TideSigProb and EncTideSigProb) for the cratering record are better favoured than other models we have tested. The trend component indicates a decreasing cratering rate ($\lambda<0$) towards the past over the past 100 Myr \citep{shoemaker98,gehrels94,mcewen97,bailer-jones11}. This suggests that either the asteroid impact rate or the preservation bias or both dominates the cratering record. Because the craters in our data sets are larger than 5\,km, the preservation bias may not be very significant over this time scale. The disruption of a single large asteroid could explain the trend in the data, as suggested by \citep{bottke07}. In addition, our models, which include the solar apex motion, can properly predict the anisotropic perihelia of LPCs without assuming a massive body in the outer Oort Cloud or an anisotropic Oort Cloud.

The EncTideSigProb, EncSigProb and TideSigProb models have Bayes factors of the
same magnitude as the SigProb model, which indicates that either the tide and encounter components are unnecessary in modelling the temporal distribution of craters, or the data cannot effectively discriminate between the models.

The stochastic component in the comet flux arising from encounters -- as represented by the term $\gamma$ -- in the EncProb and EncTideProb models can slightly increase their evidence relative to the TideProb model. We have performed a sensitivity test by changing the prior PDF over the parameters in the dynamical models and other time series models, and find only small changes of the Bayes factors. 

The asymmetrical components in the Galactic potential could, in principle, increase the time-variation of the comet flux and hence impact rate predicted by the dynamical models, by inducing larger deviations of the Sun's motion from a circular orbit and thus larger changes in the local stellar density.
It turns out that the non-axisymmetric component has relatively little impact on the predicted cometary flux, with the exception of when the Sun is in co-rotation with the spiral arms. In that case the transient resonance can produce large variations in the flux.

By including the solar apex motion, our dynamical models for anisotropic LPCs can predict reasonably well the distribution of Galactic latitude and longitude in a set of 102 dynamically new comets. In this model, the asymmetry in the distribution of Galactic latitudes caused by the Sun's current location and its motion over the past 10\,Myr (comparable with the time scale
of a comet shower). 

The two narrow peaks in the cometary perihelia at $l_c=135^\circ$ and $l_c=315^\circ$ could be caused by a handful of strong stellar encounters encountering the Sun with their encountering velocities in the direction of antapex in the HRF. On the other hand, we might also see something similar due to the periodic orbital motion about the Sun of a massive body (such as a brown dwarf) residing within the Oort cloud \citep{matese99, matese11}. However, our dynamical model, which takes into account the solar apex motion, can predict the longitudinal asymmetry without assuming the existence of such a body. In addition, the latitude distribution of LPC perihelia predicted by our simulations is consistent with the theoretical prediction, although one peak in the observed distribution is not properly predicted by our simulations. The synergy effect between the encounters and the tide cannot entirely eliminate the anisotropy induced by either the tide or the encounters.

A non-uniform distribution in the perihelion direction of encounters was found by \citet{sanchez01}, although the signal is of questionable significance due to the incompleteness, i.e.\ faint stars which high velocities being too faint after 10\,Myr for Hipparcos to have observed.

An anisotropy in the longitude of LPCs will not correspond to an anisotropy in longitudes of impacts on the Earth's surface due to the rotation of the Earth and its orbit about the Sun. Some latitude variation may be expected, despite the long-term variation in inclination and obliquity of the Earth's orbit \citep{feuvre08,werner10}. Disrupted comets generally retain their original orbital plane \citep{bottke02}, so the resulting asteroids would tend to impact in the plane perpendicular to solar apex. Yet these are all higher order effects which would be difficult to convincingly detect and relate to the solar orbit in the analysis of terrestrial impact craters.

Our modelling approach has, like any other, introduced various assumptions and approximations. We have ignored the synergy effect between the Galactic tide and stellar encounters highlighted by \citet{rickman08}. We instead simply sum the
tide-induced flux and the encounter-induced flux in the ratio $\xi$ to 1.
Because the cometary impact rate modulated by the solar motion around the
Galactic center seems to be unnecessary in order to explain the data, the
synergy effect, which is also influenced by the solar motion, may not change the
result significantly. In addition, we use a decreasing impact rate towards the past (negative trend component) to model the combined effect of preservation bias and asteroid impact rate. In modelling the angular distribution of the LPC perihelia, the sample noise in the comets injected into the observable zone prevent us from building a more robust model, especially for the longitude distribution. This problem could be resolved by calculating perturbations based on a more accurately measured Galactic tide and using an actual catalogue of encountering stars in the solar neighborhood as opposed to our stochastic model of plausible encounters. 

In common with some other studies (e.g.\ \citet{rickman08, gardner11, fouchard11, wickramasinghe08}), 
we have ignored the perturbing effect on comets from the giant planets, although we acknowledge that the giants planets could influence the predicted LPC flux in particular \citep{kaib09}. 
The planetary perturbations can also change the fraction of the inner Oort cloud comets among the injected LPCs \citep{kaib09}, which in turn could change the angular distribution of the LPC perihelia. However, these perturbations should not have a significant effect over the relatively short time scale of 10\,Myr which we use in the simulations to generate the LPC distribution.
As the main goal of our work is to study the variable effect of the solar orbit on the LPC flux and angular distribution, rather than to predict the absolute LPC flux precisely, our conclusions should not be overly affected by neglecting the giant planets in this way.

In the future, the Gaia survey allow us to detect many more recent stellar encounters down to fainter magnitude limits and larger distances than Hipparcos, thereby allowing us to extend the time scale over which we can get a complete sample of recent stellar encounters. The Gaia magnitude limit of G=20 which is low enough to cover the high velocity stars in a time scale of 10 Myr. For example, a star with absolute magnitude of 10 and a velocity of 80\,km/s in the  HRF would move 800 pc in 10 Myr and so have an apparent magnitude of 19.5. Thus Gaia will be able to observe all stars more massive than early M dwarfs (and thus essentially all relevant stars) encountering the solar system over the past 10 Myr. For more recent timescales Gaia can observe even less massive objects. Moreover, the Gaia catalogue of more massive stellar encounters (stars with absolute magnitudes larger than that of the Sun) may shed light on the study of terrestrial craters over since the beginning of the Phanerozoic era, some 550\,Myr ago. Gaia can further improve the measurement of Sun's initial conditions and the potential of the Galaxy \citep{lindegren08,koposov09}. After including planetary perturbations, this would make the simulation of cometary orbits accurate enough to trace the stellar encounter back to the time when it generated comet showers and corresponding terrestrial craters \citep{rickman12}.

\section*{Acknowledgements}
We thank Carmen Martinez, Inti Pelupessy, and Arjen van Elteren for
explaining, installing, and testing the AMUSE framework. We are indebted to Anthony Brown and Piotr A. Dybczy\'{n}ski for their useful suggestions. We are grateful to the referee, Nathan Kaib, for constructive comments which helped to improve the manuscript. This work has been carried out as part of the Gaia Research for European Astronomy Training (GREAT-ITN) network. The research leading to these results has received funding from the European Union Seventh Framework Programme ([FP7/2007-2013] under grant agreement No.\ 264895.

\bibliographystyle{mn2e}
\bibliography{comet}

\begin{thebibliography}{76}
\expandafter\ifx\csname natexlab\endcsname\relax\def\natexlab#1{#1}\fi

\bibitem[{{Alvarez} {et~al}\mbox{.}(1980){Alvarez}, {Alvarez}, {Asaro}, \&
  {Michel}}]{alvarez80}
{Alvarez} L.~W., {Alvarez} W., {Asaro} F., {Michel} H.~V., 1980, Science, 208,
  1095

\bibitem[{{Alvarez} \& {Muller}(1984)}]{alvarez84}
{Alvarez} W., {Muller} R.~A., 1984, \nat, 308, 718

\bibitem[{{Bailer-Jones}(2009)}]{bailer-jones09}
{Bailer-Jones} C.~A.~L., 2009, International Journal of Astrobiology, 8, 213

\bibitem[{{Bailer-Jones}(2011{\natexlab{a}})}]{bailer-jones11}
{Bailer-Jones} C.~A.~L., 2011{\natexlab{a}}, \mnras, 416, 1163

\bibitem[{{Bailer-Jones}(2011{\natexlab{b}})}]{2011MNRAS.418.2111B}
{Bailer-Jones} C.~A.~L., 2011{\natexlab{b}}, \mnras, 418, 2111

\bibitem[{{Bailer-Jones}(2012)}]{2012A&A...546A..89B}
{Bailer-Jones} C.~A.~L., 2012, \aap, 546, A89

\bibitem[{Bate, Mueller \& White(1971)Bate, Mueller, \& White}]{bate71}
Bate R.~R., Mueller D.~D., White J.~E., 1971, Fundamentals of astrodynamics.
  Courier Dover Publications

\bibitem[{{Biermann}, {Huebner} \& {Lust}(1983){Biermann}, {Huebner}, \&
  {Lust}}]{biermann83}
{Biermann} L., {Huebner} W.~F., {Lust} R., 1983, Proceedings of the National
  Academy of Science, 80, 5151

\bibitem[{{Binney} \& {Tremaine}(2008)}]{binney08_book}
{Binney} J., {Tremaine} S., 2008, {Galactic Dynamics: Second Edition}.
  Princeton University Press

\bibitem[{{Bogart} \& {Noerdlinger}(1982)}]{bogart82}
{Bogart} R.~S., {Noerdlinger} P.~D., 1982, \aj, 87, 911

\bibitem[{{Bottke}, {David} \& {David}(2007){Bottke}, {David}, \&
  {David}}]{bottke07}
{Bottke} W.~F., {David} V., {David} N., 2007, Nature, 449, 48,
  10.1038/nature06070

\bibitem[{{Bottke} {et~al}\mbox{.}(2002){Bottke}, {Morbidelli}, {Jedicke},
  {Petit}, {Levison}, {Michel}, \& {Metcalfe}}]{bottke02}
{Bottke} W.~F., {Morbidelli} A., {Jedicke} R., {Petit} J.-M., {Levison} H.~F.,
  {Michel} P., {Metcalfe} T.~S., 2002, \icarus, 156, 399

\bibitem[{{Brasser}, {Higuchi} \& {Kaib}(2010){Brasser}, {Higuchi}, \&
  {Kaib}}]{brasser10}
{Brasser} R., {Higuchi} A., {Kaib} N., 2010, \aap, 516, A72

\bibitem[{{Cox} \& {G{\'o}mez}(2002)}]{cox02}
{Cox} D.~P., {G{\'o}mez} G.~C., 2002, \apjs, 142, 261

\bibitem[{{Dehnen}(2000)}]{dehnen00}
{Dehnen} W., 2000, \aj, 119, 800

\bibitem[{{Dehnen} \& {Binney}(1998)}]{dehnen98b}
{Dehnen} W., {Binney} J.~J., 1998, \mnras, 298, 387

\bibitem[{{Delsemme}(1987)}]{delsemme87}
{Delsemme} A.~H., 1987, \aap, 187, 913

\bibitem[{{Dones} {et~al}\mbox{.}(2004{\natexlab{a}}){Dones}, {Levison},
  {Duncan}, \& {Weissman}}]{dones04b}
{Dones} L., {Levison} H.~F., {Duncan} M.~J., {Weissman} P.~R.,
  2004{\natexlab{a}}, Icarus, in press

\bibitem[{{Dones} {et~al}\mbox{.}(2004{\natexlab{b}}){Dones}, {Weissman},
  {Levison}, \& {Duncan}}]{dones04c}
{Dones} L., {Weissman} P.~R., {Levison} H.~F., {Duncan} M.~J.,
  2004{\natexlab{b}}, in Astronomical Society of the Pacific Conference Series,
  Vol. 323, Star Formation in the Interstellar Medium: In Honor of David
  Hollenbach, {Johnstone} D., {Adams} F.~C., {Lin} D.~N.~C., {Neufeeld} D.~A.,
  {Ostriker} E.~C., eds., p. 371

\bibitem[{{Drimmel}(2000)}]{drimmel00}
{Drimmel} R., 2000, \aap, 358, L13

\bibitem[{{Duncan}, {Quinn} \& {Tremaine}(1987){Duncan}, {Quinn}, \&
  {Tremaine}}]{duncan87}
{Duncan} M., {Quinn} T., {Tremaine} S., 1987, \aj, 94, 1330

\bibitem[{{Dybczy{\'n}ski}(2002)}]{dybczynski02}
{Dybczy{\'n}ski} P.~A., 2002, \aap, 396, 283

\bibitem[{{Dybczy{\'n}ski}(2005)}]{dybczynski05}
{Dybczy{\'n}ski} P.~A., 2005, \aap, 441, 783

\bibitem[{Emel'yanenko, Asher \& Bailey(2007)Emel'yanenko, Asher, \&
  Bailey}]{emelyanenko07}
Emel'yanenko V.~V., Asher D.~J., Bailey M.~E., 2007, \mnras, 381, 779

\bibitem[{{Feng} \& {Bailer-Jones}(2013)}]{feng13}
{Feng} F., {Bailer-Jones} C.~A.~L., 2013, \apj, 768, 152

\bibitem[{{Fouchard}(2004)}]{fouchard04}
{Fouchard} M., 2004, \mnras, 349, 347

\bibitem[{{Fouchard} {et~al}\mbox{.}(2011){Fouchard}, {Froeschl{\'e}},
  {Rickman}, \& {Valsecchi}}]{fouchard11}
{Fouchard} M., {Froeschl{\'e}} C., {Rickman} H., {Valsecchi} G.~B., 2011,
  \icarus, 214, 334

\bibitem[{{Francis}(2005)}]{francis05}
{Francis} P.~J., 2005, \apj, 635, 1348

\bibitem[{{Fujii} {et~al}\mbox{.}(2007){Fujii}, {Iwasawa}, {Funato}, \&
  {Makino}}]{fujii07}
{Fujii} M., {Iwasawa} M., {Funato} Y., {Makino} J., 2007, \pasj, 59, 1095

\bibitem[{{Garc{\'{\i}}a-S{\'a}nchez}
  {et~al}\mbox{.}(1999){Garc{\'{\i}}a-S{\'a}nchez}, {Preston}, {Jones},
  {Weissman}, {Lestrade}, {Latham}, \& {Stefanik}}]{sanchez99}
{Garc{\'{\i}}a-S{\'a}nchez} J., {Preston} R.~A., {Jones} D.~L., {Weissman}
  P.~R., {Lestrade} J.-F., {Latham} D.~W., {Stefanik} R.~P., 1999, \aj, 117,
  1042

\bibitem[{{Garc{\'{\i}}a-S{\'a}nchez}
  {et~al}\mbox{.}(2001){Garc{\'{\i}}a-S{\'a}nchez}, {Weissman}, {Preston},
  {Jones}, {Lestrade}, {Latham}, {Stefanik}, \& {Paredes}}]{sanchez01}
{Garc{\'{\i}}a-S{\'a}nchez} J., {Weissman} P.~R., {Preston} R.~A., {Jones}
  D.~L., {Lestrade} J.-F., {Latham} D.~W., {Stefanik} R.~P., {Paredes} J.~M.,
  2001, \aap, 379, 634

\bibitem[{{Gardner} {et~al}\mbox{.}(2011){Gardner}, {Nurmi}, {Flynn}, \&
  {Mikkola}}]{gardner11}
{Gardner} E., {Nurmi} P., {Flynn} C., {Mikkola} S., 2011, \mnras, 411, 947

\bibitem[{{Gehrels}, {Matthews} \& {Schumann}(1994){Gehrels}, {Matthews}, \&
  {Schumann}}]{gehrels94}
{Gehrels} T., {Matthews} M.~S., {Schumann} A.~M., eds., 1994, {Hazards due to
  comets and asteroids}

\bibitem[{{Grieve} \& {Pesonen}(1996)}]{1996EM&P...72..357G}
{Grieve} R.~A.~F., {Pesonen} L.~J., 1996, Earth Moon and Planets, 72, 357

\bibitem[{{Heisler}, {Tremaine} \& {Alcock}(1987){Heisler}, {Tremaine}, \&
  {Alcock}}]{heisler87}
{Heisler} J., {Tremaine} S., {Alcock} C., 1987, \icarus, 70, 269

\bibitem[{{Hildebrand} {et~al}\mbox{.}(1991){Hildebrand}, {Penfield}, {Kring},
  {Pilkington}, {Camargo Z.}, {Jacobsen}, \& {Boynton}}]{1991Geo....19..867H}
{Hildebrand} A.~R., {Penfield} G.~T., {Kring} D.~A., {Pilkington} M., {Camargo
  Z.} A., {Jacobsen} S.~B., {Boynton} W.~V., 1991, Geology, 19, 867

\bibitem[{{Hills}(1981)}]{hills81}
{Hills} J.~G., 1981, \aj, 86, 1730

\bibitem[{{Jetsu} \& {Pelt}(2000)}]{2000A&A...353..409J}
{Jetsu} L., {Pelt} J., 2000, \aap, 353, 409

\bibitem[{{Kaib} \& {Quinn}(2009)}]{kaib09}
{Kaib} N.~A., {Quinn} T., 2009, Science, 325, 1234

\bibitem[{Kaib, Quinn \& Brasser(2011)Kaib, Quinn, \& Brasser}]{kaib11b}
Kaib N.~A., Quinn T., Brasser R., 2011, The Astronomical Journal, 141, 3

\bibitem[{{Kaib}, {Ro{\v s}kar} \& {Quinn}(2011){Kaib}, {Ro{\v s}kar}, \&
  {Quinn}}]{kaib11}
{Kaib} N.~A., {Ro{\v s}kar} R., {Quinn} T., 2011, \icarus, 215, 491

\bibitem[{Kass \& Raftery(1995)}]{kass95}
Kass R.~E., Raftery A.~E., 1995, Journal of the American Statistical
  Association, 90, 773–795

\bibitem[{{Khanna} \& {Sharma}(1983)}]{khanna83}
{Khanna} M., {Sharma} S.~D., 1983, \pasj, 35, 559

\bibitem[{{Koposov}, {Rix} \& {Hogg}(2010){Koposov}, {Rix}, \&
  {Hogg}}]{koposov09}
{Koposov} S.~E., {Rix} H.-W., {Hogg} D.~W., 2010, \apj, 712, 260

\bibitem[{{Le Feuvre} \& {Wieczorek}(2008)}]{feuvre08}
{Le Feuvre} M., {Wieczorek} M.~A., 2008, \icarus, 197, 291

\bibitem[{{Levison} {et~al}\mbox{.}(2010){Levison}, {Duncan}, {Brasser}, \&
  {Kaufmann}}]{levison10}
{Levison} H.~F., {Duncan} M.~J., {Brasser} R., {Kaufmann} D.~E., 2010, Science,
  329, 187

\bibitem[{{Lindegren} {et~al}\mbox{.}(2008){Lindegren}, {Babusiaux},
  {Bailer-Jones}, {Bastian}, {Brown}, {Cropper}, {H{\o}g}, {Jordi}, {Katz},
  {van Leeuwen}, {Luri}, {Mignard}, {de Bruijne}, \& {Prusti}}]{lindegren08}
{Lindegren} L. {et~al.}, 2008, in IAU Symposium, Vol. 248, IAU Symposium, {Jin}
  W.~J., {Platais} I., {Perryman} M.~A.~C., eds., pp. 217--223

\bibitem[{Luhman(2014)}]{luhman14}
Luhman K.~L., 2014, The Astrophysical Journal, 781, 4

\bibitem[{{Majaess}, {Turner} \& {Lane}(2009){Majaess}, {Turner}, \&
  {Lane}}]{majaess09}
{Majaess} D.~J., {Turner} D.~G., {Lane} D.~J., 2009, \mnras, 398, 263

\bibitem[{{Marsden} \& {Williams}(2008)}]{marsden08}
{Marsden} B.~G., {Williams} G.~V., 2008, {Catalogue of Cometary Orbits 2008.
  17th edition.}

\bibitem[{{Martos} {et~al}\mbox{.}(2004){Martos}, {Ya{\~n}ez}, {Hernandez},
  {Moreno}, \& {Pichardo}}]{martos04}
{Martos} M., {Ya{\~n}ez} M., {Hernandez} X., {Moreno} E., {Pichardo} B., 2004,
  Journal of Korean Astronomical Society, 37, 199

\bibitem[{{Matese} \& {Lissauer}(2002)}]{matese02}
{Matese} J.~J., {Lissauer} J.~J., 2002, \icarus, 157, 228

\bibitem[{{Matese}, {Whitman} \& {Whitmire}(1999){Matese}, {Whitman}, \&
  {Whitmire}}]{matese99}
{Matese} J.~J., {Whitman} P.~G., {Whitmire} D.~P., 1999, \icarus, 141, 354

\bibitem[{{Matese} \& {Whitmire}(2011)}]{matese11}
{Matese} J.~J., {Whitmire} D.~P., 2011, \icarus, 211, 926

\bibitem[{{McEwen}, {Moore} \& {Shoemaker}(1997){McEwen}, {Moore}, \&
  {Shoemaker}}]{mcewen97}
{McEwen} A.~S., {Moore} J.~M., {Shoemaker} E.~M., 1997, \jgr, 102, 9231

\bibitem[{Melott \& Bambach(2011)}]{melott11}
Melott A.~L., Bambach R.~K., 2011, Paleobiology, 37, 383

\bibitem[{{Oort}(1950)}]{oort50}
{Oort} J.~H., 1950, \bain, 11, 91

\bibitem[{{Pelupessy}, {J{\"a}nes} \& {Portegies Zwart}(2012){Pelupessy},
  {J{\"a}nes}, \& {Portegies Zwart}}]{pelupessy12}
{Pelupessy} F.~I., {J{\"a}nes} J., {Portegies Zwart} S., 2012, \na, 17, 711

\bibitem[{{Pelupessy} {et~al}\mbox{.}(2013){Pelupessy}, {van Elteren}, {de
  Vries}, {McMillan}, {Drost}, \& {Portegies Zwart}}]{pelupessy13}
{Pelupessy} F.~I., {van Elteren} A., {de Vries} N., {McMillan} S.~L.~W.,
  {Drost} N., {Portegies Zwart} S.~F., 2013, \aap, 557, A84

\bibitem[{{Portegies Zwart} {et~al}\mbox{.}(2013){Portegies Zwart}, {McMillan},
  {van Elteren}, {Pelupessy}, \& {de Vries}}]{portegies13}
{Portegies Zwart} S., {McMillan} S.~L.~W., {van Elteren} E., {Pelupessy} I.,
  {de Vries} N., 2013, Computer Physics Communications, 183, 456

\bibitem[{{Raup} \& {Sepkoski}(1984)}]{raup84}
{Raup} D.~M., {Sepkoski} J.~J., 1984, Proceedings of the National Academy of
  Science, 81, 801

\bibitem[{{Rickman}(1976)}]{rickman76}
{Rickman} H., 1976, Bulletin of the Astronomical Institutes of Czechoslovakia,
  27, 92

\bibitem[{{Rickman} {et~al}\mbox{.}(2008){Rickman}, {Fouchard},
  {Froeschl{\'e}}, \& {Valsecchi}}]{rickman08}
{Rickman} H., {Fouchard} M., {Froeschl{\'e}} C., {Valsecchi} G.~B., 2008,
  Celestial Mechanics and Dynamical Astronomy, 102, 111

\bibitem[{{Rickman} {et~al}\mbox{.}(2012){Rickman}, {Fouchard},
  {Froeschl{\'e}}, \& {Valsecchi}}]{rickman12}
{Rickman} H., {Fouchard} M., {Froeschl{\'e}} C., {Valsecchi} G.~B., 2012,
  \planss, 73, 124

\bibitem[{{Rickman} {et~al}\mbox{.}(2005){Rickman}, {Fouchard}, {Valsecchi}, \&
  {Froeschl{\'e}}}]{rickman05}
{Rickman} H., {Fouchard} M., {Valsecchi} G.~B., {Froeschl{\'e}} C., 2005, Earth
  Moon and Planets, 97, 411

\bibitem[{{Rohde} \& {Muller}(2005)}]{rohde05}
{Rohde} R.~A., {Muller} R.~A., 2005, \nat, 434, 208

\bibitem[{{Sch{\"o}nrich}(2012)}]{schoenrich12}
{Sch{\"o}nrich} R., 2012, \mnras, 427, 274

\bibitem[{{Sch{\"o}nrich}, {Binney} \& {Dehnen}(2010){Sch{\"o}nrich}, {Binney},
  \& {Dehnen}}]{schoenrich10}
{Sch{\"o}nrich} R., {Binney} J., {Dehnen} W., 2010, \mnras, 403, 1829

\bibitem[{{Shoemaker}(1998)}]{shoemaker98}
{Shoemaker} E.~M., 1998, \jrasc, 92, 297

\bibitem[{{Wainscoat} {et~al}\mbox{.}(1992){Wainscoat}, {Cohen}, {Volk},
  {Walker}, \& {Schwartz}}]{wainscoat92}
{Wainscoat} R.~J., {Cohen} M., {Volk} K., {Walker} H.~J., {Schwartz} D.~E.,
  1992, \apjs, 83, 111

\bibitem[{{Weissman}(2007)}]{weissman07}
{Weissman} P.~R., 2007, in IAU Symposium, Vol. 236, IAU Symposium, {Valsecchi}
  G.~B., {Vokrouhlick{\'y}} D., {Milani} A., eds., pp. 441--450

\bibitem[{{Werner} \& {Medvedev}(2010)}]{werner10}
{Werner} S.~C., {Medvedev} S., 2010, Earth and Planetary Science Letters, 295,
  147

\bibitem[{{Wickramasinghe} \& {Napier}(2008)}]{wickramasinghe08}
{Wickramasinghe} J.~T., {Napier} W.~M., 2008, \mnras, 387, 153

\bibitem[{Wiegert \& Tremaine(1999)}]{wiegert99}
Wiegert P., Tremaine S., 1999, Icarus, 137, 84

\bibitem[{{Wisdom} \& {Holman}(1991)}]{wisdom91}
{Wisdom} J., {Holman} M., 1991, \aj, 102, 1528

\bibitem[{{Yabushita}(1996)}]{1996MNRAS.279..727Y}
{Yabushita} S., 1996, \mnras, 279, 727

\end{thebibliography}
\label{lastpage}
\end{document}